\documentclass[a4paper,11pt]{article}
\pdfoutput=1
\usepackage{jheppub}

\usepackage{amssymb,amsmath}
\usepackage[normalem]{ulem}
\usepackage[utf8x]{inputenc}
\usepackage{slashed}
\usepackage{graphicx}
\usepackage{here}
\usepackage{color}
\usepackage{csquotes} %fix backwards quotes
\usepackage{comment}
\usepackage{mathrsfs}
\usepackage{float}
\usepackage{ascmac}
\usepackage{multirow}
\usepackage{longtable}
\usepackage{enumitem}

\usepackage[hang,small,bf]{caption}
\usepackage[subrefformat=parens]{subcaption}
\usepackage{hyperref}

\usepackage{ulem}
\usepackage{xcolor}

\DeclareRobustCommand{\eraseblue}{\bgroup\markoverwith{\textcolor{blue}{\rule[.5ex]{2pt}{0.4pt}}}\ULon}
\DeclareRobustCommand{\erasered}{\bgroup\markoverwith{\textcolor{red}{\rule[.5ex]{2pt}{0.4pt}}}\ULon}

\newcommand{\greekii}{I\hspace{-.1em}I}

\makeatletter
\newcommand*\rel@kern[1]{\kern#1\dimexpr\macc@kerna}
\newcommand*\widebar[1]{%
  \begingroup
  \def\mathaccent##1##2{%
    \rel@kern{0.8}%
    \overline{\rel@kern{-0.8}\macc@nucleus\rel@kern{0.2}}%
    \rel@kern{-0.2}%
  }%
  \macc@depth\@ne
  \let\math@bgroup\@empty \let\math@egroup\macc@set@skewchar
  \mathsurround\z@ \frozen@everymath{\mathgroup\macc@group\relax}%
  \macc@set@skewchar\relax
  \let\mathaccentV\macc@nested@a
  \macc@nested@a\relax111{#1}%
  \endgroup
}
\makeatother

\numberwithin{equation}{section}

\preprint{
\begin{minipage}{5cm}
\small
\flushright
EPHOU-25-013\\KYUSHU-HET-331
\end{minipage}}

\title{Stringy Constraints on Modular Flavor Models}

\author{Keiya Ishiguro$^{1}$,} 
\author{Takafumi Kai$^{2}$,}
\author{Tatsuo Kobayashi$^{3}$, and} 
\author{Hajime Otsuka$^{2}$} 
\affiliation{
$^1$SMBC Nikko Securities Inc. Quantitative Analytics \& Developments.,  1-5-1, Marunouchi, Chiyoda-Ku, Tokyo, Japan.\\
$^2$Department of Physics, Kyushu University, 744 Motooka, Nishi-ku, Fukuoka 819-0395, Japan\\
$^3$Department of Physics, Hokkaido University, Sapporo 060-0810, Japan
}
\emailAdd{keyspire@ruri.waseda.jp}
\emailAdd{kai.takafumi@phys.kyushu-u.ac.jp}
\emailAdd{kobayashi@particle.sci.hokudai.ac.jp}
\emailAdd{otsuka.hajime@phys.kyushu-u.ac.jp}

\abstract{
We investigate stringy constraints on moduli spaces in modular flavor models by analyzing moduli-dependent threshold corrections in heterotic string vacua. 
While moduli play a crucial role in determining the flavor structure of fermions predicted by modular flavor models, the parameter space in which their vacuum expectation values are allowed has not been fully explored. 
In this work, within the framework of perturbative heterotic string theory on toroidal orbifolds, we derive constraints on the moduli space and study their systematic behavior. 
We characterize the stringy constraints in terms of the dilaton, the beta-function coefficients, and the ratio between a complex-structure modulus and a K\"ahler modulus. 
It is found that the large value of the modulus controlling the flavor structure, i.e., $\tau\simeq i \infty$, is inconsistent within perturbative string theory, and the self-dual point $\tau=i$ is also disfavored in the large volume regime of toroidal backgrounds. 
In addition, we discuss the phenomenological implications of these stringy constraints.  
}
\makeatletter
\gdef\@fpheader{}
\makeatother

\begin{document}

\maketitle

%%%%%%%%%%%%%%%%%%%%%%%%%%%%%%%%%%%%%%%%%%%%%%%%%%%%%%%%%%%%
\section{Introduction}
\label{sec:intro}
%%%%%%%%%%%%%%%%%%%%%%%%%%%%%%%%%%%%%%%%%%%%%%%%%%%%%%%%%%%%

Particle physics has important mysteries concerning flavor structures of quarks and leptons, which are associated with a generation number, mass hierarchies, mixing angles and CP phases.
The mysteries indicate an existence of a more fundamental theory and hold potential to search for new physics beyond the Standard Model.
Non-Abelian discrete flavor symmetry is one of the fundamental ideas to explain the flavor structure (see, e.g., Refs. \cite{Altarelli_2010, Ishimori_2010, King_2013, king2015modelsneutrinomassmixing,Kobayashi:2022moq}).
The flavor symmetry based on finite discrete groups, e.g. $S_3$, $A_4$, $S_4$, $A_5$ and other groups, produces interesting mixing patterns and they are tested to high precision.
Moreover, since the modular group includes the finite discrete groups \cite{deAdelhartToorop:2011re} and modular invariance in flavor models enables fewer parameters to reveal the flavor structure than the traditional flavor symmetry, modular flavor symmetry is considered to be a more fundamental symmetry in the flavor sector \cite{Feruglio:2017spp}. (See Refs.~\cite{Kobayashi:2023zzc,Ding:2023htn} for reviews).

\medskip

In the modular flavor models, considering that modular transformations act on a complex modulus $\tau$, we construct moduli-dependent Yukawa couplings, which are modular forms transforming non-trivially under finite modular groups such as $S_3, A_4, S_4$, and $A_5$ \cite{Kobayashi:2018vbk,Feruglio:2017spp,Penedo:2018nmg,Novichkov:2018nkm}.
A vacuum expectation value (VEV) of the modulus is a source of flavor symmetry breaking and the modulus VEV determines the masses and flavor mixing of both quarks and leptons, CP violation phases including Dirac and Majorana phases.
In the context of higher-dimensional theory, the modulus $\tau$ originates from geometric degrees of freedom regarding an extra-dimensional space, which are called complex-structure moduli and K\"{a}hler moduli \cite{Strominger:1985it, CANDELAS1988458, DIXON199027, Cremades:2004wa}.
Thus, the modular flavor symmetry is also associated with 10D supergravity and string theory, and the relation between the flavor symmetry and the higher-dimensional theory has increased the significance of top-down model building with the modular symmetry and the extra-dimensional space, 
e.g. in heterotic orbifolds \cite{Ferrara:1989qb,Lerche:1989cs,Lauer:1990tm,Nilles:2021glx} and type IIB magnetized D-brane models \cite{ Kobayashi_2017, Kobayashi:2018rad,Kobayashi:2018bff,Ohki:2020bpo,Kikuchi:2020frp,Kikuchi:2020nxn,
Kikuchi:2021ogn,Almumin:2021fbk}.

\medskip

While the modulus plays an important role in predictions of the flavor structure in the modular flavor model, parameter spaces where the modulus VEV should be allowed remain unexplored. 
In the context of type IIB flux compactifications, the modulus VEV controlling the flavor structure of fermions is constrained by the tadpole cancellation condition, and it is statistically fixed at fixed points of $PSL(2,\mathbb{Z})$ modular symmetry on $T^6/(\mathbb{Z}_2\times \mathbb{Z}_2)$ orbifold~\cite{Ishiguro:2020tmo}, or $\bar{\Gamma}_0(N)$ modular symmetry on the other toroidal orbifolds~\cite{Ishiguro:2023wwf,Ishiguro:2025pwa}.
In this study, we propose stringy constraints on the moduli spaces by taking account of one-loop threshold corrections to gauge couplings in heterotic string vacua \cite{kaplunovsky:1992, Dixon:1990pc, Kaplunovsky_1995}.
When considering low-energy effective theories of string theory, the modular flavor models are necessary to follow perturbative theories concerning string vacua.
Since the moduli-dependent threshold corrections must be smaller than tree-level gauge couplings, we can derive the constraints on the moduli spaces in order to guarantee the perturbative theories of heterotic string vacua.
This study classifies the threshold corrections with target space duality symmetries in Ref. \cite{BAILIN:1994} and uses them to explore the stringy constraints.
In addition, the constraint that radii in heterotic compactifications cannot be arbitrarily large is a well-known fact, as shown in Refs. \cite{ANTONIADIS1990377, IBANEZ1991251, BARBIERI199479, FLORAKIS20171, ANGELANTONJ2019496}.
Thus, by utilizing the perturbative criteria, which have been discussed extensively in the GUT vs. string scale problem and also the decompactification problem, our study explores quantitative insights into a specific $\mathcal{O}(1)$ region with respect to the moduli space.

\medskip

This paper is organized as follows.
In Sec. \ref{sec:modular_flavor}, we first review the modular flavor models in the context of the higher-dimensional theory.
Then, wavefunctions in $d$-dimensional compact space have the geometric information of compact space and follow non-trivial representations under the modular transformation regarding deformations of the compact space.
This representation is associated with the flavor symmetry in the four-dimensional(4D) effective theory.
In Sec. \ref{sec:threshold_corection}, we focus on the moduli-dependent threshold corrections to gauge couplings at energy scales below a string scale.
Here, we review the general formula of string loop threshold corrections in heterotic string theory on toroidal orbifolds.
In Sec. \ref{sec:stringy_constraints}, we present the explicit moduli-dependent threshold corrections on $T^6/\mathbb{Z}_{4}$ orbifold and $T^6/\mathbb{Z}_{6-\rm \greekii}$ orbifold.
Considering these explicit threshold corrections, we use the target space duality symmetries to classify the threshold corrections on each orbifold and perform a systematic analysis for the stringy constraints.
In Sec. \ref{sec:analysis_of_stringy_constraints}, we explore the explicit stringy constraints concerning the moduli spaces by using the threshold corrections which are classified systematically.
First, a duality group $PSL_T(2, \mathbb{Z}) \times PSL_U(2, \mathbb{Z})$ is analyzed through a dependence of beta-function coefficients.
Second, we perform the same analysis for the other duality symmetries.
Note that we focus solely on cases where exists the contributions of the threshold corrections, in other words, instanton numbers for two $E_8$ gauge groups are not equal in the context of M-theory.
In Sec. \ref{sec:phenomenological_aspects}, we show that the stringy constraints which are constructed in the previous sections can constrain the moduli space on modular flavor models.
Especially, large imaginary parts of moduli cannot be allowed from the viewpoint of the perturbative theories.
When considering a case of large volume regarding the K\"{a}hler modulus, almost all of the moduli space cannot be allowed and the allowed region concentrates on a fixed point of the duality symmetry.
Sec. \ref{sec:conclusion} is devoted to the conclusions.

%%%%%%%%%%%%%%%%%%%%%%%%%%%%%%%%%%%%%%%%%%%%%%%%%%%%%%%%%%%%
\section{Modular flavor models from higher-dimensional theory}
\label{sec:modular_flavor}
%%%%%%%%%%%%%%%%%%%%%%%%%%%%%%%%%%%%%%%%%%%%%%%%%%%%%%%%%%%%

In this section, we briefly review a model with modular symmetry from the viewpoint of higher-dimensional theory, following Ref.~\cite{Kikuchi:2022txy}.

Let us consider a higher-dimensional theory on a direct product of 4D spacetime and $d$-dimensional compact space with modular symmetry. We represent bosonic fields $\Phi(x,y)$ including scalars or vectors in $(4+d)$-dimensional spacetime and fermionic fields $\Psi(x,y)$, where the coordinates $x$ and $y$ respectively represent 4D spacetime and $d$-dimensional compact space. 
Through Kaluza-Klein reduction, $\Phi(x,y)$ and $\Psi(x,y)$ can be decomposed as
\begin{align}
    \Phi(x,y) &= \sum_i \phi_i(x)\varphi_i(y) + (\mathrm{KK~modes})\,,
    \nonumber\\
    \Psi(x,y) &= \sum_i \psi_i(x)\chi_i(y) + (\mathrm{KK~modes})\,,
\end{align}
where we focus on massless modes and the index $i$ denotes their degeneracy, corresponding to the flavor index in the 4D low-energy effective field theory. 
Note that $\{\phi(x), \psi(x)\}$ correspond to 4D massless fields, and $\{\varphi(y), \chi(y)\}$ are wavefunctions in $d$-dimensional compact space. 
When the $d$-dimensional compact space has the modular symmetry, e.g, $SL(2,\mathbb{Z})$ modular symmetry, the wavefunctions $\{\varphi(y), \chi(y)\}$ are described by a certain modular forms depending on a modulus $\tau$. For instance, twisted modes on heterotic $T^2/\mathbb{Z}_3$ orbifold are described by representations of $T'$~\cite{Ferrara:1989qb,Lerche:1989cs,Lauer:1990tm}, and other representations appear in other string models, e.g., type IIB magnetized D-brane models~\cite{Kobayashi:2018rad,Kobayashi:2018bff,Ohki:2020bpo,Kikuchi:2020frp,Kikuchi:2020nxn,
Kikuchi:2021ogn,Almumin:2021fbk} and heterotic string on Calabi-Yau threefolds with standard embedding~\cite{Ishiguro:2020nuf,Ishiguro:2021ccl,Ishiguro:2024xph}. 

To show the presence of modular symmetry in the 4D low-energy effective action, let us consider the canonical kinetic term of $\Phi$ in a higher-dimensional theory:
\begin{align}
    \partial_M \Phi^\ast \partial^M \Phi
\end{align}
with the index $M$ runs the coordinates $x$ and $y$. When the $d$-dimensional compact space enjoys the modular symmetry, e.g., $PSL(2,\mathbb{Z})$, zero-mode wavefunctions $\varphi_i(y)$ will be described by $d_{k_i}$-dimensional modular forms of $\bar{\Gamma}(N)$ with weight $k_i$, where $\bar{\Gamma}(N)$ is the principal congruence subgroup of $PSL(2,\mathbb{Z})$:
\begin{align}
\begin{aligned}
\bar{\Gamma}(N)= \left \{ 
\begin{pmatrix}
a & b  \\
c & d  
\end{pmatrix} \in PSL(2,\mathbb{Z})~ ,
~~
\begin{pmatrix}
a & b  \\
c & d  
\end{pmatrix} =
\begin{pmatrix}
1 & 0  \\
0 & 1  
\end{pmatrix} ~~({\rm mod}~N) \right \}
\end{aligned} .
\end{align}
It means that under the modular transformation of $\tau$:
\begin{align}
\label{eq:tautrf}
\tau \to \gamma \tau = \frac{a \tau +b}{c \tau +d}\,,
\end{align}
with $\{a,b,c,d\}$ being integers satisfying $ad-bc=1$, the wavefunction $\varphi_i(y)$ transforms as
\begin{align}
\label{eq:trfvarphi}
    \varphi_i(y)\rightarrow \rho(\gamma)_{ij} (c\tau+d)^{k_i} \varphi_j(y),
\end{align}
where $\rho(\gamma)$ is the unitary matrix. 
To extract the 4D modular symmetric action, we normalize the internal wavefunction $\varphi_i(y)$ as
\begin{align}
    \int d^d y \sqrt{g} |\varphi_i(y)|^2 = (2{\rm Im}\tau)^{- k_i},
\end{align}
where $g$ denotes the determinant of the $d$-dimensional metric. 
Note that this normalization is modular invariant under Eqs.~\eqref{eq:tautrf} and~\eqref{eq:trfvarphi} because $\int d^d y \sqrt{g}$ is the modular invariant quantity. 

By adopting the above normalization of $\varphi_i(y)$, one can obtain the modular invariant K\"ahler potential of matter fields $\phi_i(x)$:
\begin{align}
K =  \sum_i \frac{1}{(2{\rm Im}(\tau ))^{k_i}}|\phi_i(x)|^2.
\label{eq:Kahler}
\end{align}
Similarly, one can discuss the holomorphic Yukawa couplings of matter fields, which can be derived by overlap integrals of three matter wavefunctions:
\begin{align}
Y_{ij\ell}(\tau) \simeq \int d^dy~\sqrt{g}~ \chi_{i}(y) ~\chi_j(y) ~\varphi^*_\ell(y)\,,
\end{align}
up to the normalization factor. From the modular weights of matters, one can extract the modular weight of the above holomorphic Yukawa coupling, i.e., $k_i+k_j-k_\ell$, and the modular transformation is given by Eq.~\eqref{eq:trfvarphi}:
\begin{align}
    Y(\tau) \rightarrow (c\tau+d)^{k_i+k_j-k_\ell}\rho_Y(\gamma) Y(\tau)\,,
\end{align}
with $\rho_Y(\gamma)$ being the tensor product of representation matrices of matters. 
This feature can also be understood by the operator production of matter fields, as demonstrated in toroidal backgrounds~\cite{Cremades:2004wa,Abe:2009dr} and more general curved backgrounds~\cite{Honda:2018sjy}. Indeed, two point couplings of matter wavefunctions $\chi_i(y)\chi_j(y)$ can be expanded as
\begin{align}
\label{eq:OPE}    
\chi_i(y)\chi_j(y)=\sum_\ell Y_{ij\ell}\chi_\ell(y) + (\mathrm{KK~modes}),
\end{align}
due to the fact that a complete set of wavefunctions can be constructed by including all the KK wavefunctions. 
The modular weight of the coefficient $Y_{ij\ell}$ appearing in the right hand side of Eq.~\eqref{eq:OPE} can be determined by the modular weights of matters. 

So far, vacuum expectation values of moduli fields determining the flavor structure of massless modes are fixed by hand to fit them with the observed value of fermion masses and mixing angles in a bottom-up approach. On the other hand, on a top-down approach, the moduli values are determined by a certain stabilization mechanism in the framework of modular invariant theory such as type IIB flux compactifications~\cite{Ishiguro:2020tmo,Ishiguro:2022pde,Ishiguro:2023wwf}, gaugino condensations in heterotic string theory~\cite{Novichkov:2022wvg}, generating moduli potential due to the matter-modular mixings~\cite{Knapp-Perez:2023nty}, and radiative moduli stabilization~\cite{Kobayashi:2023spx,Higaki:2024jdk}.\footnote{Spontaneous CP violation was also discussed in Refs.~\cite{Kobayashi:2020uaj,Ishiguro:2020nuf,Higaki:2024pql}.} 
In the context of type IIB flux compactifications, an amount of flux quanta is constrained by the cancellation condition of D3-brane charges. It is found that the small tadpole charge favored by the tadpole cancellation condition leads to the stabilization of moduli values at fixed points of $PSL(2,\mathbb{Z})$ modular symmetry on $T^6/(\mathbb{Z}_2\times \mathbb{Z}_2)$ orbifold~\cite{Ishiguro:2020tmo}, or $\bar{\Gamma}_0(N)$ modular symmetry on the other toroidal orbifolds~\cite{Ishiguro:2023wwf,Ishiguro:2025pwa}. 
These fixed points of modular symmetry lead to various phenomenologically interesting features~\cite{Novichkov:2018ovf,Novichkov:2018yse,Novichkov:2018nkm,Ding:2019gof,Okada:2019uoy,King:2019vhv,Okada:2020rjb,Okada:2020ukr,Okada:2020brs,Feruglio:2021dte,Kobayashi:2021pav,Kobayashi:2022jvy} including the dark matter (DM) stability \cite{Kobayashi:2021ajl} since the residual symmetries remain at the fixed points. 

It is quite important to reveal stringy constraints on moduli values in other corner of string compactifications. 
In the latter part of this paper, we study moduli-dependent threshold corrections to gauge couplings on the basis of heterotic string theory on toroidal orbifolds. 
Then, we extract constraints on moduli values taking into account the validity of threshold corrections and the realization of realistic values of gauge couplings.

%%%%%%%%%%%%%%%%%%%%%%%%%%%%%%%%%%%%%%%%%%%%%%%%%%%%%%%%%%%%
\section{Moduli-dependent threshold corrections to gauge couplings}
\label{sec:threshold_corection}
%%%%%%%%%%%%%%%%%%%%%%%%%%%%%%%%%%%%%%%%%%%%%%%%%%%%%%%%%%%%

At energies below a string scale, a gauge group derived from the gauge structure of heterotic string theory has a product structure $G = \prod_a G_a$.
All gauge groups $G_a$ have related tree-level gauge couplings as follows:
\begin{align}
    \frac{1}{g_a^2} = \frac{k_a}{g_{\rm string}^2}\,,
\end{align}
where $g_{\rm string}$ is a string coupling parameter and $k_a$ is a level of Ka\v{c}-Moody current algebra giving rise to $G_a$.
In addition, these gauge couplings are modified by finite threshold corrections due to loop effects regarding charged heavy particles.
At one-loop level of string theory, effective gauge couplings are given by 
\begin{align}
    \frac{16 \pi^2}{g_a^2(\mu)} = k_a ~ \frac{16 \pi^2}{g_{\rm string}^2} + b_a ~  \text{log} \frac{M_{\rm string}^2}{\mu^2} + \Delta_a,
    \label{eq:couplings_with_corrections}
\end{align}
where $\mu$ is a momentum scale at which the gauge couplings are measured and $b_a$ are coefficients of one-loop beta functions on a low-energy effective theory \cite{kaplunovsky:1992, Dixon:1990pc, Kaplunovsky_1995}.
Here, $\Delta_a$ are one-loop string threshold corrections to the gauge coupling $g_a^{-2}$ and the general formula of $\Delta_a$ for orbifold compactification is discussed in Ref. \cite{kaplunovsky:1992} and valid for any four-dimensional tachyon-free vacuum of the heterotic string.
Besides, other studies \cite{Mayr:1993mq, BAILIN:1994} discuss the threshold corrections on more specific orbifolds.

Since we would like to derive stringy constraints on moduli, we focus on moduli-dependent string loop threshold corrections.
The threshold corrections are determined by twisted sectors $(h, g)$ of the orbifold, which has $\mathcal{N} = 2$ supersymmetry.
Explicitly, the formula of $\Delta_a$ is denoted as
\begin{align}
    \Delta_a = \int_{\mathcal{F}} \frac{d^2 \tau}{\tau_2} \sum_{(h, g)} b_a^{(h, g)} \mathcal{Z}_{(h, g)} (\tau, \bar{\tau}) - b_a^{\mathcal{N} = 2} \int_{\mathcal{F}} \frac{d^2 \tau}{\tau_2},
    \label{eq:pre_threshold_correction}
\end{align}
where $\mathcal{Z}_{(h, g)}$ are the moduli dependent parts of the partition functions for the $\mathcal{N} = 2$ twisted sectors $(h, g)$, $b_a^{(h, g)}$ is the contribution of $(h, g)$ sector to the coefficient of the one-loop beta functions, $b_a^{\mathcal{N} = 2}$ is the coefficient of the beta function for all $\mathcal{N} = 2$ twisted sectors and $\mathcal{F}$ is a fundamental region for world sheet modular group $PSL(2, \mathbb{Z})$ \cite{BAILIN:1994}:
\begin{align}
    \mathcal{F} = \left\{ - \frac{1}{2} \leq \tau_1 \leq 0, \quad |\tau| \geq 1 \right\} ~ \cup ~ \left\{ 0 < \tau_1 < \frac{1}{2}, \quad |\tau| > 1 \right\},
\end{align}
where $\tau = \tau_1 + i \tau_2$.
For convenience, considering the discussion in \cite{Mayr:1993mq}, we rewrite Eq. \eqref{eq:pre_threshold_correction} in terms of a subset of $\mathcal{N} = 2$ twisted sectors $(h_0, g_0)$ with the integration over an enlarged region $\tilde{\mathcal{F}}$ as follows:
\begin{align}
    \Delta_a = \sum_{(h_0, g_0)} b_a^{(h_0, g_0)} \int_{\tilde{\mathcal{F}}} \frac{d^2 \tau}{\tau_2} \mathcal{Z}_{(h_0, g_0)} (\tau, \bar{\tau}) - b_a^{\mathcal{N} = 2} \int_{\mathcal{F}} \frac{d^2 \tau}{\tau_2}.
    \label{eq:threshold_correction}
\end{align}
Here, a set of twisted sectors can be obtained from $PSL(2, \mathbb{Z})$ transformations of the single twisted sector $(h_0, g_0)$ and the enlarged region $\tilde{\mathcal{F}}$ is also generated by $PSL(2, \mathbb{Z})$ transformations of the fundamental region $\mathcal{F}$.
Then, the partition function $\mathcal{Z}_{(h_0, g_0)}$ is invariant under a congruence subgroup $\bar{\Gamma}_0(n)$ which is defined as
\begin{align}
    \bar{\Gamma}_0(n) \equiv \left\{\begin{pmatrix}
        a & b \\
        c & d
    \end{pmatrix} \in PSL(2, \mathbb{Z}) \middle| ~ c \equiv 0 \quad (\text{mod}~n)
    \right\}.
\end{align}
Then, $\bar{\Gamma}^0(n)$ is also defined as follows:
\begin{align}
    \bar{\Gamma}^0(n) \equiv \left\{\begin{pmatrix}
        a & b \\
        c & d
    \end{pmatrix} \in PSL(2, \mathbb{Z}) \middle| ~ b \equiv 0 \quad (\text{mod}~n)
    \right\}.
\end{align}
Therefore, the enlarged region can be indicated by generators ${\cal S}$ and $\cal T$ of $PSL(2, \mathbb{Z})$.
In addition, $\Gamma_0(n)$ (or $\Gamma^0(n)$) is defined in the same way as $\bar{\Gamma}_0(n)$ (or $\bar{\Gamma}^0(n)$) except that $SL(2, \mathbb{Z})$ is used instead of $PSL(2, \mathbb{Z})$.
In the case of $\bar{\Gamma}_0(2)$, the enlarged region is denoted as
\begin{align}
    \tilde{\mathcal{F}} = \{I, {\cal S}, {\cal S}{\cal T}\}\mathcal{F}.
\end{align}
In the case of $\bar{\Gamma}_0(3)$, the enlarged region is denoted as
\begin{align}
    \tilde{\mathcal{F}} = \{I,{\cal S}, {\cal ST}, {\cal S}{\cal T}^2 \}\mathcal{F}.
\end{align}

%%%%%%%%%%%%%%%%%%%%%%%%%%%%%%%%%%%%%%%%%%%%%%%%%%%%%%%%%%%%
\section{Classification of threshold corrections}
\label{sec:stringy_constraints}
%%%%%%%%%%%%%%%%%%%%%%%%%%%%%%%%%%%%%%%%%%%%%%%%%%%%%%%%%%%%

In this section, we classify the moduli-dependent threshold corrections and derive a condition to ensure the perturbative validity of heterotic string vacua.
Since the string loop threshold corrections $\Delta_a$ for non-decomposable $\mathbb{Z}_{N}$ orbifolds are known to be calculated in Ref.~\cite{Dixon:1990pc, BAILIN:1994}, we prepare the explicit forms of the threshold corrections by using the modular symmetries, which are necessary to construct inequalities such that the tree-level gauge coupling is larger than the threshold corrections.

%%%%%%%%%%%%%%%%%%%%%%%%%%%%%%%%%%%%%%%%%%%%%%%%%%%%%%%%%%%%
\subsection{Basic threshold corrections}
\label{sec:basic}
%%%%%%%%%%%%%%%%%%%%%%%%%%%%%%%%%%%%%%%%%%%%%%%%%%%%%%%%%%%%

As discussed in Ref. \cite{Dixon:1990pc}, basic formulas of the string loop threshold correction on each toroidal orbifold can be denoted as follows:
\begin{align}
    \Delta_a = - b_a^{\mathcal{N} = 2} ~ \text{ln} \left[ \frac{8 \pi}{3 \sqrt{3}} e^{1 - \gamma_{\rm E}} ~ T_2 |\eta(T)|^4 ~ U_2 |\eta(U)|^4 \right].
    \label{eq:basic_delta}
\end{align}
Let us discuss a target space modular symmetry on Eq. \eqref{eq:basic_delta}.
The generators $\cal S$ and $\cal T$ regarding $SL(2, \mathbb{Z})$ are denoted as follows:
\begin{align}
    {\cal S} = \begin{pmatrix}
        0 & -1 \\
        1 & 0
    \end{pmatrix}, \qquad
    {\cal T} = \begin{pmatrix}
        1 & 1 \\
        0 & 1
    \end{pmatrix},
\end{align}
and, when considering a linear fractional transformation of $SL(2, \mathbb{Z})$ modular group, which is defined by
\begin{align}
    \gamma : \tau \rightarrow \gamma \cdot \tau = \frac{a \tau + b}{c \tau + d}, \qquad 
    \left( 
    \gamma = 
    \begin{pmatrix}
        a & b \\
        c & d
    \end{pmatrix} \in SL(2, \mathbb{Z}) \right)
\end{align} 
the linear fractional transformations that correspond to the generators $\cal S$ and $\cal T$ can be described as follows:
\begin{align}
    {\cal S} : \tau \rightarrow 
    \begin{pmatrix}
        0 & -1 \\
        1 & 0
    \end{pmatrix} \cdot \tau = - \frac{1}{\tau}, \qquad 
    {\cal T} : \tau \rightarrow 
    \begin{pmatrix}
        1 & 1 \\
        0 & 1
    \end{pmatrix} \cdot \tau = \tau + 1.
\end{align}
As shown by the properties of the linear fractional transformations, the transformation identifies $\pm \gamma$ as $\gamma$.
Therefore, we need to deal with the $PSL(2, \mathbb{Z})$ modular group when considering the linear fractional transformation based on $SL(2, \mathbb{Z})$.

Since the modular transformation properties with respect to the $\eta$-function can be represented by the following relations based on the linear fractional transformations:
\begin{align}
    \eta \left( - \frac{1}{\tau} \right) = \sqrt{- i \tau} ~ \eta (\tau), \qquad \eta(\tau + 1) = e^{\pi i /12}\eta(\tau),
\end{align}
the symmetries of the threshold corrections must be described by the subgroup of $PSL(2, \mathbb{Z})$.
In addition, the symmetry of $PSL(2, \mathbb{Z})$ appears obviously on the function $\text{Im}\tau |\eta(\tau)|^4$.
Therefore, taking account of these transformations on the $\eta$-function, one can find that Eq. \eqref{eq:basic_delta} displays the target space modular symmetry $PSL_T(2, \mathbb{Z}) \times PSL_U(2, \mathbb{Z})$.

As shown in Ref. \cite{BAILIN:1994}, there are not only the above threshold correction Eq. \eqref{eq:basic_delta}, but also other examples whose target space modular symmetries are different from $PSL_T(2,\mathbb{Z}) \times PSL_U (2,\mathbb{Z})$.
For example, threshold corrections on $T^6/\mathbb{Z}_4$ possesses the target space modular symmetry $\bar{\Gamma}_{T}^0(2) \times (\bar{\Gamma}_{U})_0(2)$, and threshold corrections on $T^6/\mathbb{Z}_{6-\rm \greekii}$ possesses the target space modular symmetry $\bar{\Gamma}_{T}^0(3) \times (\bar{\Gamma}_{U})_0(3)$ and $PSL_{\hat{T}}(2, \mathbb{Z}) \times PSL_{\hat{U}}(2, \mathbb{Z})$.
Especially, it is noteworthy that phenomenologically interesting models on $T^6/\mathbb{Z}_{6-\rm \greekii}$ orbifolds are constructed as discussed in Refs. \cite{Kobayashi_2005, Buchm_ller_2007, Lebedev_2007, Lebedev_2008}.
To classify the threshold corrections based on various toroidal orbifolds, we focus on the two examples of $T^6/\mathbb{Z}_4$ and $T^6/\mathbb{Z}_{6-\rm \greekii}$ orbifolds as benchmarks.

%%%%%%%%%%%%%%%%%%%%%%%%%%%%%%%%%%%%%%%%%%%%%%%%%%%%%%%%%%%%
\subsection{$T^6/\mathbb{Z}_{4}$ orbifold with $SU(2) \times SU(4) \times SO(5)$ root lattice}
\label{sec:SU(2)SU(4)SO(5)}
%%%%%%%%%%%%%%%%%%%%%%%%%%%%%%%%%%%%%%%%%%%%%%%%%%%%%%%%%%%%

Firstly, we focus on the threshold corrections of $T^6/\mathbb{Z}_{4}$ orbifold with $SU(2) \times SU(4) \times SO(5)$ root lattice, where a $\mathbb{Z}_4$ twist is given by $\theta = \text{exp}\left[ \frac{2 \pi i}{4} (1, 1, -2) \right]$:
\begin{align}
    \begin{aligned}
        \Delta_a = &- b_a^{(I, \theta^2)} ~ \text{ln} \left( \frac{8 \pi}{3 \sqrt{3}} e^{1 - \gamma_{\rm E}} ~ T_2 |\eta(T)|^4 U_2 |\eta(2U)|^4 \right) \\
        &- b_a^{(I, \theta^2)} ~ \text{ln} \left( \frac{8 \pi}{3 \sqrt{3}} e^{1 - \gamma_{\rm E}} ~ T_2 \left| \eta \left( \frac{T}{2} \right) \right|^4 U_2 |\eta(U)|^4 \right),
    \end{aligned}
\end{align}
where $b_a^{(I, \theta^2)}$ is the contribution of $(I, \theta^2)$ sector to the coefficient of the one-loop beta functions, $\gamma_{\rm E}$ is Euler-Mascheroni constant and $\eta(\tau)$ is Dedekind eta-function.
Here, $T$ and $U$ are respectively K\"{a}hler modulus and complex-structure modulus associated with a fixed plane in $(I, \theta^2)$ and they are defined as $T = T_1 + iT_2$ and $U = U_1 + i U_2$.

Since a complex twist is given by $\theta$ on $T^6/\mathbb{Z}_{4}$ orbifold with $SU(2) \times SU(4) \times SO(5)$ root lattice, $\mathcal{N} = 2$ orbit is denoted as
\begin{align}
    \mathcal{O} = \{ (I, \theta^2), (\theta^2, I), (\theta^2, \theta^2) \}.
\end{align}
This orbit is rewritten by a fundamental orbit $\mathcal{O}_0$ which consists of $(I, \theta^2)$ in the same way as the enlarged region $\tilde{\mathcal{F}} = \{I,{\cal S}, {\cal S}{\cal T} \}\mathcal{F}$.
Considering each contribution generated from $b_a^{(I, \theta^2)}$ and a volume factor $V_2 = 2$ \cite{Erler_1993}, the value of $b_a^{\mathcal{N} = 2}$ can be related to $b_a^{(I, \theta^2)}$ as follows:
\begin{align}
    b_a^{\mathcal{N} = 2} = b_a^{(I, \theta^2)} \left( 1 + \frac{1}{2} + \frac{1}{2} \right) = 2 b_a^{(I, \theta^2)}.
    \label{eq:theta2twistZ4}
\end{align}
By using these relations Eq. \eqref{eq:theta2twistZ4}, we can write the threshold correction $\Delta_a$ with $b_a^{\mathcal{N} = 2}$
\footnote{
In this paper, for simplicity, we focus on only the coefficient of the one-loop beta function even though there are some numerical constants associated with Green-Schwarz function \cite{Kaplunovsky_1995}.
},
\begin{align}
    \begin{aligned}
        \Delta_a = - b_a^{\mathcal{N} = 2} ~ \text{ln} \left[ \frac{8 \pi}{3 \sqrt{3}} e^{1 - \gamma_{\rm E}} ~ T_2 |\eta(T)|^2 \left| \eta \left( \frac{T}{2} \right) \right|^2 ~ U_2 |\eta(U)|^2| \eta(2U)|^2 \right].
    \end{aligned}
    \label{eq:explicit_DeltaZ4}
\end{align}

Let us discuss a target space modular symmetry on Eq. \eqref{eq:explicit_DeltaZ4}.
The function $|\eta(T/2)|^2$ is not invariant under the $\cal T$ transformation as follows:
\begin{align}
    {\cal T} : \left| \eta \left( \frac{T}{2} \right) \right|^2 \rightarrow \left| \eta \left( \frac{T + 1}{2} \right) \right|^2 = \left| \eta \left( \frac{T}{2} + \frac{1}{2} \right) \right|^2,
\end{align}
and likewise, the function $|\eta(2 U)|^2$ is not invariant under the $\cal STS$ transformation as follows:
\begin{align}
    {\cal STS} : |\eta(2 U)|^2 \rightarrow \left| \eta \left( \frac{-2 U}{U - 1} \right) \right|^2 = \left| \eta \left( \frac{-1 \cdot (2U)}{ \frac{1}{2} \cdot (2U) - 1} \right) \right|^2.
\end{align}
Here, we consider the actions ${\cal T}^2$ and $ {\cal S}{\cal T}^2{\cal S}$ on the functions $|\eta(T/2)|^2$ and  $|\eta(2 U)|^2$ respectively,
\begin{align}
    \begin{aligned}
        {\cal T}^2 &: \left| \eta \left( \frac{T}{2} \right) \right|^2 \rightarrow \left| \eta \left( \frac{T + 2}{2} \right) \right|^2 = \left| \eta \left( \frac{T}{2} \right) \right|^2, \\
        {\cal S}{\cal T}^2{\cal S} &: |\eta(2 U)|^2 \rightarrow \left| \eta \left( \frac{-2 U}{2U - 1} \right) \right|^2 = |2U - 1| |\eta(2 U)|^2.
    \end{aligned}
\end{align}
A subgroup that is generated by $\cal STS$ and ${\cal T}^2$ is known as $\Gamma^0(2)$, and likewise, a subgroup that is generated by $\cal T$ and ${\cal S}{\cal T}^2{\cal S}$ is known as $\Gamma_0(2)$.
Therefore, taking account of the linear fractional transformations on the $\eta$-function, one can find that Eq. \eqref{eq:explicit_DeltaZ4} displays the target space modular symmetry $\bar{\Gamma}_{T}^0(2) \times (\bar{\Gamma}_{U})_0(2)$.

%%%%%%%%%%%%%%%%%%%%%%%%%%%%%%%%%%%%%%%%%%%%%%%%%%%%%%%%%%%%
\subsection{$T^6/\mathbb{Z}_{6-\rm \greekii}$ orbifold with $SU(2) \times SU(3) \times SO(7)$ root lattice}
\label{}
%%%%%%%%%%%%%%%%%%%%%%%%%%%%%%%%%%%%%%%%%%%%%%%%%%%%%%%%%%%%

Next, we focus on the threshold corrections of $T^6/\mathbb{Z}_{6-\rm \greekii}$ orbifold with $SU(2) \times SU(3) \times SO(7)$ root lattice, where a $\mathbb{Z}_{6-\rm \greekii}$ twist is given by $\theta = \text{exp}\left[ \frac{2 \pi i}{6} (2, 1, -3) \right]$.
The explicit form of $\Delta_a$ is described as
\begin{align}
    \begin{aligned}
        \Delta_a = &- 2 b_a^{(I, \theta^2)} ~ \text{ln} \left( \frac{8 \pi}{3 \sqrt{3}} e^{1 - \gamma_{\rm E}} ~ T_2 |\eta(T)|^4 U_2 |\eta(U)|^4 \right) \\
        &- 2 b_a^{(I, \theta^2)} ~ \text{ln} \left( \frac{8 \pi}{3 \sqrt{3}} e^{1 - \gamma_{\rm E}} ~ T_2 \left| \eta \left( \frac{T}{3} \right) \right|^4 U_2 |\eta(3U)|^4 \right) \\
        &- \hat{b}_a ~ \text{ln} \left( \frac{8 \pi}{3 \sqrt{3}} e^{1 - \gamma_{\rm E}} ~ \hat{T}_2 |\eta ( \hat{T})|^4 \hat{U}_2 |\eta(\hat{U})|^4 \right),
    \end{aligned}
\end{align}
where $b_a^{(I, \theta^2)}$ is also the contribution of $(I, \theta^2)$ sector, and $T$ and $U$ are respectively K\"{a}hler modulus and complex-structure modulus associated with a fixed plane in $(I, \theta^2)$ and they are defined as $T = T_1 + iT_2$ and $U = U_1 + i U_2$.
Here, $\hat{b}_a$ is the contribution of $(I, \theta^3)$ sector to the coefficient of the one-loop beta functions, $\hat{T}$ and $\hat{U}$ are respectively the K\"{a}hler modulus and the complex-structure modulus associated with a fixed plane in $(I, \theta^3)$ and they are defined as $\hat{T} = \hat{T}_1 + i \hat{T}_2$ and $\hat{U} = \hat{U}_1 + i \hat{U}_2$.

Since a complex twist is given by $\theta$ on $T^6/\mathbb{Z}_{6-\rm \greekii}$ orbifold with $SU(2) \times SU(3) \times SO(7)$ root lattice, $\mathcal{N} = 2$ orbit is denoted as
\begin{align}
    \mathcal{O} = \{ (I, \theta^2), (\theta^2, I), (\theta^2, \theta^2), (\theta^2, \theta^4), (I, \theta^4), (\theta^4, I), (\theta^4, \theta^4), (\theta^4, \theta^2) \}.
\end{align}
This orbit is rewritten by a fundamental orbit $\mathcal{O}_0$ which consists of $(I, \theta^2)$ and $(I, \theta^4)$ in the same way as the enlarged region $\tilde{\mathcal{F}} = \{I, {\cal S}, {\cal ST}, {\cal S}{\cal T}^2 \}\mathcal{F}$.
Considering each contribution generated from $b_a^{(I, \theta^2)}$ and a volume factor $V_2 = 3$, the value of $b_a^{\mathcal{N} = 2}$ can be related to $b_a^{(I, \theta^2)}$ as follows:
\begin{align}
    b_a^{\mathcal{N} = 2} = 2 b_a^{(I, \theta^2)} \left( 1 + \frac{1}{3} + \frac{1}{3} + \frac{1}{3} \right) = 4 b_a^{(I, \theta^2)}.
    \label{eq:theta2twistZ6}
\end{align}
By using these relations Eq. \eqref{eq:theta2twistZ6}, we can write the threshold correction $\Delta_a$ with $b_a^{\mathcal{N} = 2}$
\begin{align}
    \begin{aligned}
        \Delta_a = &- b_a^{\mathcal{N} = 2} ~ \text{ln} \left[ \frac{8 \pi}{3 \sqrt{3}} e^{1 - \gamma_{\rm E}} ~ T_2 |\eta(T)|^2 \left| \eta \left( \frac{T}{3} \right) \right|^2 ~ U_2 |\eta(U)|^2| \eta(3U)|^2 \right] \\
        &- \hat{b}_a ~ \text{ln} \left[ \frac{8 \pi}{3 \sqrt{3}} e^{1 - \gamma_{\rm E}} ~ \hat{T}_2 |\eta ( \hat{T})|^4 \hat{U}_2 |\eta(\hat{U})|^4 \right].
    \end{aligned}
    \label{eq:explicit_DeltaZ6}
\end{align}
Then, by taking account of the same discussions in Sec. \ref{sec:SU(2)SU(4)SO(5)}, Eq. \eqref{eq:explicit_DeltaZ6} displays the target space modular symmetry $\bar{\Gamma}_{T}^0(3) \times (\bar{\Gamma}_{U})_0(3)$ and $PSL_{\hat{T}}(2, \mathbb{Z}) \times PSL_{\hat{U}}(2, \mathbb{Z})$.

%%%%%%%%%%%%%%%%%%%%%%%%%%%%%%%%%%%%%%%%%%%%%%%%%%%%%%%%%%%%
\subsection{General structure of threshold corrections}
\label{sec:general}
%%%%%%%%%%%%%%%%%%%%%%%%%%%%%%%%%%%%%%%%%%%%%%%%%%%%%%%%%%%%

In this section, we classify the threshold corrections by using the target space modular symmetry $PSL(2, \mathbb{Z})$ and $\bar{\Gamma}_{T}^0(n) \times (\bar{\Gamma}_{U})_0(n)$.
For the specific forms of the threshold corrections, they can be roughly divided into the following five groups \cite{Dixon:1990pc, BAILIN:1994}:
\begin{enumerate}
    \item $PSL_T(2, \mathbb{Z}) \times PSL_U(2, \mathbb{Z})$
    \begin{align}
        \begin{aligned}
            \Delta_a = - b_a^{\mathcal{N} = 2} ~ \text{ln} \left[ \frac{8 \pi}{3 \sqrt{3}} e^{1 - \gamma_{\rm E}} ~ T_2 |\eta(T)|^4 ~ U_2 |\eta(U)|^4 \right],
        \end{aligned}
        \label{eq:classifyPSL}
    \end{align}

    \item $\bar{\Gamma}_{T}^0(2) \times (\bar{\Gamma}_{U})_0(2)$
    \begin{align}
        \begin{aligned}
            \Delta_a = - b_a^{\mathcal{N} = 2} ~ \text{ln} \left[ \frac{8 \pi}{3 \sqrt{3}} e^{1 - \gamma_{\rm E}} ~ T_2 |\eta(T)|^2 \left| \eta \left( \frac{T}{2} \right) \right|^2 ~ U_2 |\eta(U)|^2| \eta(2U)|^2 \right],
        \end{aligned}
        \label{eq:classifyGamma22}
    \end{align}

    \item $\bar{\Gamma}_{T}^0(2) \times PSL(2, \mathbb{Z})_U$
    \begin{align}
        \begin{aligned}
            \Delta_a = - b_a^{\mathcal{N} = 2} ~ \text{ln} \left[ \frac{8 \pi}{3 \sqrt{3}} e^{1 - \gamma_{\rm E}} ~ \frac{1}{4^{2/3}} ~ T_2 \left| \eta \left( \frac{T}{2} \right) \right|^4 ~ U_2 |\eta(U)|^4 \right],
        \end{aligned}
        \label{eq:classifyGamma2SL}
    \end{align}

    \item $\bar{\Gamma}_{T}^0(3) \times (\bar{\Gamma}_{U})_0(3)$ and $PSL_{\hat{T}}(2, \mathbb{Z}) \times PSL_{\hat{U}}(2, \mathbb{Z})$
    \begin{align}
        \begin{aligned}
            \Delta_a = &- b_a^{\mathcal{N} = 2} ~ \text{ln} \left[ \frac{8 \pi}{3 \sqrt{3}} e^{1 - \gamma_{\rm E}} ~ T_2 |\eta(T)|^2 \left| \eta \left( \frac{T}{3} \right) \right|^2 ~ U_2 |\eta(U)|^2| \eta(3U)|^2 \right] \\
            &- b_a^{\mathcal{N} = 2} ~ \text{ln} \left[ \frac{8 \pi}{3 \sqrt{3}} e^{1 - \gamma_{\rm E}} ~ \hat{T}_2 |\eta ( \hat{T})|^4 \hat{U}_2 |\eta(\hat{U})|^4 \right],
        \end{aligned}
        \label{eq:classifyGamma3}
    \end{align}

    \item $\bar{\Gamma}_{T}^0(3) \times (\bar{\Gamma}_{U})_0(3)$ and $\bar{\Gamma}^0_{\hat{T}}(2)$
    \begin{align}
        \begin{aligned}
            \Delta_a = &- b_a^{\mathcal{N} = 2} ~ \text{ln} \left[ \frac{8 \pi}{3 \sqrt{3}} e^{1 - \gamma_{\rm E}} ~ T_2 |\eta(T)|^2 \left| \eta \left( \frac{T}{3} \right) \right|^2 ~ U_2 |\eta(U)|^2| \eta(3U)|^2 \right] \\
            &- b_a^{\mathcal{N} = 2} ~ \text{ln} \left[ \frac{8 \pi}{3 \sqrt{3}} e^{1 - \gamma_{\rm E}} ~ \frac{1}{4^{2/3}} ~ \hat{T}_2 \left| \eta \left( \frac{\hat{T}}{2} \right) \right|^4 \frac{\sqrt{3}}{2} |\eta(\omega)|^4 \right],
        \end{aligned}
        \label{eq:classifyGamma32}
    \end{align}
\end{enumerate}
where $\omega=- \frac{1}{2} + i \frac{\sqrt{3}}{2}$ is the fixed point of $PSL(2,\mathbb{Z})$, and for simplicity, we choose $\hat{b}_a = b_a^{\mathcal{N} = 2}$. 
Although these groups cannot classify all forms of the threshold corrections on toroidal orbifolds, their classification broadly captures the characteristics of threshold corrections, as all other forms of the function can be characterized by combinations of $PSL(2, \mathbb{Z})$, $\bar{\Gamma}_0(n)$ and $\bar{\Gamma}^0(n)$
\footnote{
The target space duality symmetry $\bar{\Gamma}_T^0 (2)$ in Ref. \cite{BAILIN:1994} is included in the group of $\bar{\Gamma}_{T}^0(2) \times PSL(2, \mathbb{Z})_U$.
}.
For example, the threshold correction of the group (1) $PSL_T(2, \mathbb{Z}) \times PSL_U(2, \mathbb{Z})$ is the simplest form of all groups and it is explored in detail in Ref. \cite{Dixon:1990pc}.
Moreover, according to Ref. \cite{BAILIN:1994}, this form appears in the $(I, \theta^3)$ sector of the threshold corrections concerning $T^6/\mathbb{Z}_{6-\rm \greekii}$ orbifold with $SU(2) \times SU(3) \times SO(7)$ root lattice and $T^6/\mathbb{Z}_{6-\rm \greekii}$ orbifold with $SU(3) \times SO(8)$ root lattice.
The orbifolds appearing in Ref. \cite{BAILIN:1994} are classified by using the groups as follows:
\begin{description}
    \item[Group~(2)] $\bar{\Gamma}_{T}^0(2) \times (\bar{\Gamma}_{U})_0(2)$
    
        $T^6/\mathbb{Z}_4$ orbifold with $SU(2) \times SU(4) \times SO(5)$ root lattice
        
        $T^6/\mathbb{Z}_{8-\rm \greekii}$ orbifold with $SU(2) \times SO(10)$ root lattice

    \item[Group~(3)] $\bar{\Gamma}_{T}^0(2) \times PSL(2, \mathbb{Z})_U$
    
        $T^6/\mathbb{Z}_4$ orbifold with $SU(4) \times SU(4)$ root lattice

        $T^6/\mathbb{Z}_{12-\rm I}$ orbifold with $E_6$ root lattice

    \item[Group~(4)] $\bar{\Gamma}_{T}^0(3) \times (\bar{\Gamma}_{U})_0(3)$ and $PSL_{\hat{T}}(2, \mathbb{Z}) \times PSL_{\hat{U}}(2, \mathbb{Z})$

        $T^6/\mathbb{Z}_{6-\rm \greekii}$ orbifold with $SU(2) \times SU(3) \times SO(7)$ root lattice

    \item[Group~(5)] $\bar{\Gamma}_{T}^0(3) \times (\bar{\Gamma}_{U})_0(3)$ and $\bar{\Gamma}^0_{\hat{T}}(2)$
    
        $T^6/\mathbb{Z}_{6-\rm \greekii}$ orbifold with $SU(2) \times SU(6)$ root lattice
    
\end{description}
In addition, $\bar{\Gamma}_{T}^0(3) \times \bar{\Gamma}_{U + 2}^0(3)$ and $PSL_{\hat{T}}(2, \mathbb{Z}) \times PSL_{\hat{U}}(2, \mathbb{Z})$ appears in $T^6/\mathbb{Z}_{6-\rm \greekii}$ orbifold with $SU(3) \times SO(8)$ root lattice \cite{BAILIN:1994}. 
However, the threshold correction is similar to the group (1) in the sense that it is invariant with respect to the permutation of $T$ and $U + 2$.

%%%%%%%%%%%%%%%%%%%%%%%%%%%%%%%%%%%%%%%%%%%%%%%%%%%%%%%%%%%%
\section{Stringy constraints for moduli}
\label{sec:analysis_of_stringy_constraints}
%%%%%%%%%%%%%%%%%%%%%%%%%%%%%%%%%%%%%%%%%%%%%%%%%%%%%%%%%%%%

For $\mathcal{N} = 1$ supersymmetric vacua of the heterotic string, the gauge couplings are calculated directly from the string-theoretical amplitudes and $g_{\rm string}^{-2}$ depends solely on a dilaton \cite{WITTEN1985151}.
At the tree level, the gauge couplings can be defined as
\begin{align}
    f_a \equiv \frac{1}{g_a^2} - \frac{i \theta_a}{8 \pi^2} = k_a (i \bar{S}),
    \label{eq:holomorphiccoupling}
\end{align}
where $S$ is the bosonic part of the axio-dilaton \cite{Kaplunovsky_1995, Kaplunovsky_1994}.
Here, to be consistent with the notation of the K\"{a}hler modulus, this notation of axio-dilaton is chosen in Eq. \eqref{eq:holomorphiccoupling}.
Note that, in this paper, we discuss $g_{\rm string}^{-2} = \text{Im}S + \mathcal{O}(1)$ as the effective gauge coupling.

For an effective quantum field theory of the heterotic string vacua, Wilsonian gauge couplings $f_a$ are holomorphic functions of the chiral superfields and depend solely on the one-loop level of the perturbative theory.
To discuss these features for the Wilsonian gauge couplings, the supersymmetric cutoff is purely perturbative \cite{Mayr_1994, Kaplunovsky_1995}.
Therefore, we assume that the gauge couplings are weak enough at the string scale.
Taking account of the weak gauge couplings, we can derive the following inequalities by using Eq. \eqref{eq:couplings_with_corrections},
\begin{align}
    16 \pi^2 k_a \text{Im} S > |\Delta_a|,
    \label{eq:inequality1}
\end{align}
where we choose $\mu^2 \sim M_{\rm string}^2$.
$\Delta_a$ have the dependence of the K\"ahler modulus and the complex-structure modulus.
These moduli correspond to a geometrical internal degree of freedom in the toroidal orbifolds, and it is associated with one of the degrees of freedom that can predict flavor observables in modular invariant models of flavor physics.
Hence, the inequality Eq. \eqref{eq:inequality1} can constrain the parameter space of the modulus of the modular flavor models in the regions whether the perturbative description is guaranteed or not.

Note that, within the framework of M-theory, the threshold corrections vanish in the four-dimensional low-energy effective theory if instanton numbers for two $E_8$ gauge groups are equal \cite{STIEBERGER1999109}.
Therefore, our study focuses on the case that the contribution of threshold corrections appears in the effective theories.

%%%%%%%%%%%%%%%%%%%%%%%%%%%%%%%%%%%%%%%%%%%%%%%%%%%%%%%%%%%%
\subsection{Constraints for complex-structure modulus and K\"{a}hler modulus}
\label{}
%%%%%%%%%%%%%%%%%%%%%%%%%%%%%%%%%%%%%%%%%%%%%%%%%%%%%%%%%%%%

\begin{figure}[H]
\centering
\includegraphics[width = 0.8 \linewidth]{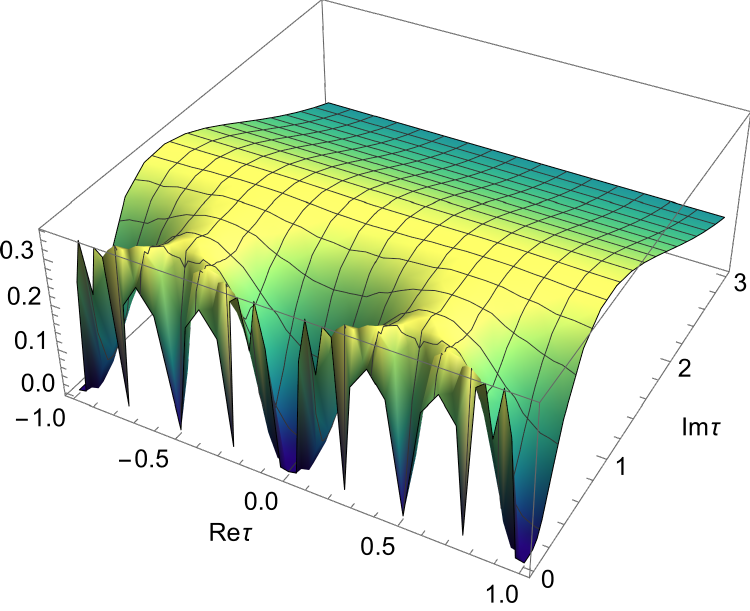}
\caption{The shape of $\text{Im} \tau |\eta(\tau)|^4$.}
\label{fig:shape_of_function}
\end{figure}

Specifically, we discuss how this inequality limits the complex-structure modulus and the K\"{a}hler modulus.
By using Eq. \eqref{eq:classifyPSL} and Eq. \eqref{eq:inequality1}, the inequality can be written as follows:
\begin{align}
    \begin{aligned}
        16 \pi^2 k_a \text{Im} S > \left| - b_a^{\mathcal{N} = 2} ~ \text{ln} \left[ \frac{8 \pi}{3 \sqrt{3}} e^{1 - \gamma_{\rm E}} ~ T_2 |\eta(T)|^4 ~ U_2 |\eta(U)|^4 \right] \right|.
    \end{aligned}
    \label{eq:inequality2}
\end{align}
Next, we evaluate a function of $\text{Im} \tau |\eta(\tau)|^4$ with $\tau=T$ or $U$.
In the right-hand side (RHS) of Eq. \eqref{eq:inequality2}, the complex-structure modulus $U$ and the K\"{a}hler moduli $T$ remain as the degrees of freedom, and the constraints change due to the values of the moduli.
In general, the function of $\text{Im} \tau |\eta(\tau)|^4$ has the maximum value, and the shape is shown in Fig. \ref{fig:shape_of_function}.

\noindent
Therefore, by choosing the moduli at the points where $T_2 |\eta(T)|^4 ~ U_2 |\eta(U)|^4$ is maximized, we can obtain a minimum bound on the inequality in the moduli space.
Then, we can find $U \sim - 0.500 + 0.866 i$ and $T \sim - 0.500 + 0.866 i$ numerically as these points.
In addition, by assuming minimal supersymmetric Standard Model (MSSM) in the low-energy effective theory, the value of $\text{Im}S$ can be found as $\text{Im}S \sim 2$ from the value of MSSM gauge couplings around the grand unification scale.
For the level of Ka\v{c}-Moody current algebra, we choose $k_a = 1$ in this work for SU(3) and SU(2) gauge couplings.
For the coefficient of the one-loop beta function, we use the typical value $|b_a^{\mathcal{N} = 2}| = \mathcal{O}(10)$ and check the behavior of the stringy constraints depending on the beta-function coefficients.
Considering these numerical constants, the RHS of Eq. \eqref{eq:inequality2} can be calculated as follows:
\begin{align}
    - b_a^{\mathcal{N} = 2} ~ \text{ln} \left[ \frac{8 \pi}{3 \sqrt{3}} e^{1 - \gamma_{\rm E}} ~ T_2 |\eta(T)|^4 ~ U_2 |\eta(U)|^4 \right] > 0.680 \dots,
    \label{eq:RHS}
\end{align}
where $b_a^{\mathcal{N} = 2}$ is chosen as 10.
It is found that the minimum value of RHS satisfies Eq. \eqref{eq:inequality2} and the RHS with the typical values $b_a^{\mathcal{N} = 2} = \mathcal{O}(10)$ also satisfies Eq. \eqref{eq:inequality2}
\footnote{
For the case of $b_a^{\mathcal{N} = 2} < 0$, the inequality can be denoted as
\begin{align}
    \begin{aligned}
        - 16 \pi^2 k_a \text{Im} S < - b_a^{\mathcal{N} = 2} ~ \text{ln} \left[ \frac{8 \pi}{3 \sqrt{3}} e^{1 - \gamma_{\rm E}} ~ T_2 |\eta(T)|^4 ~ U_2 |\eta(U)|^4 \right].
    \end{aligned}
\end{align}
This provides the same results as the case of $b_a^{\mathcal{N} = 2} > 0$.
Note that $\text{ln} \left[ \frac{8 \pi}{3 \sqrt{3}} e^{1 - \gamma_{\rm E}} ~ T_2 |\eta(T)|^4 ~ U_2 |\eta(U)|^4 \right]$ is negative on the upper half plane regarding the complex-structure modulus and the K\"{a}hler modulus.
}.

Next, we focus on the behavior of the RHS for Eq. \eqref{eq:inequality2} when the moduli, the beta-function coefficient and the dilaton change respectively.
Firstly, we prepare the approximate form of the Dedekind eta function on the specific moduli space in order to give analytical constraints for the moduli.
The explicit form of the Dedekind eta function can be denoted as $q$-series,
\begin{align}
    \eta (\tau) = \sum_{n = -\infty}^{\infty} (-1)^n q^{\frac{(6n-1)^2}{24}} = q^{\frac{1}{24}} [1 - q + \dots], \qquad (q \equiv e^{2 \pi i \tau}).
\end{align}
If we restrict the parameter space of the complex-structure modulus to $\text{Im} U > 1$, the Dedekind eta function satisfies $\eta(U) \sim e^{\frac{2 \pi i U}{24}}$ approximately.
Therefore, we can rewrite the function of $\text{ln} \left[ U_2 |\eta(U)|^4 \right] $ as follows:
\begin{align}
    \text{ln} \left[ U_2 |\eta(U)|^4 \right] \sim - \frac{\pi}{3} U_2 + \text{ln} U_2.
    \label{eq:approx}
\end{align}
The same result is obtained for the K\"{a}hler modulus.

In what follows, we study constraints on $U$ when we fix $T$ or relate $T$ to $U$.
We can derive the same results for $T$ when we fix $U$ or relate $U$ to $T$.

\paragraph{Fixing the modulus to $\omega$} \mbox{}

\begin{figure}[H]
\centering
\includegraphics[width = 0.6 \linewidth]{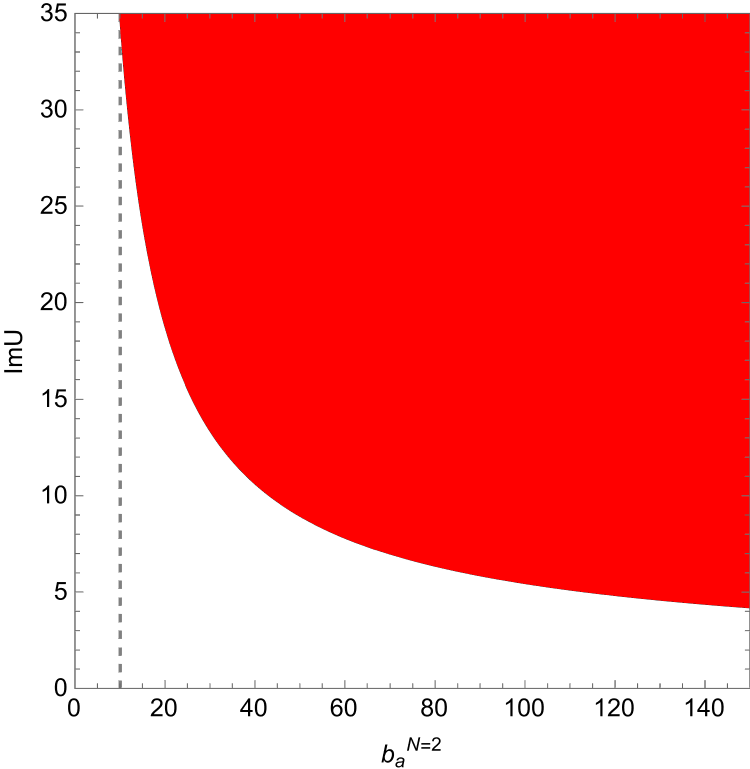}
\caption{This is the restriction imposed on the imaginary part of the complex-structure modulus while fixing the K\"{a}hler modulus to $T = - \frac{1}{2} + i \frac{\sqrt{3}}{2}$.
The red region is the parameter space where the perturbative description cannot be guaranteed.}
\label{fig:stcon}
\end{figure}

As the approximate function Eq. \eqref{eq:approx} can be obtained, we examine the restrictions on the moduli space arising from the inequality Eq. \eqref{eq:inequality2}.
Here, we vary only the complex-structure modulus while fixing the K\"{a}hler modulus to $T = - \frac{1}{2} + i \frac{\sqrt{3}}{2}$, since the complex-structure modulus and the K\"{a}hler modulus follow the same functional form $\text{Im} \tau |\eta(\tau)|^4$.
By using Eq. \eqref{eq:inequality2} and Eq. \eqref{eq:approx}, we can obtain the explicit constraint for the imaginary part of the complex-structure modulus:
\begin{align}
    \frac{\pi}{3} U_2 - \text{ln} U_2 < \text{ln} \left[ \frac{8 \pi}{3 \sqrt{3}} e^{1 - \gamma_{\rm E}} ~ \frac{\sqrt{3}}{2} |\eta(\omega)|^4 \right] + \frac{16 \pi^2 \text{Im} S}{b_a^{\mathcal{N} = 2}} = A,
    \label{eq:approx_inequlity}
\end{align}
where we choose $T = \omega = - \frac{1}{2} + i \frac{\sqrt{3}}{2}$ as the K\"{a}hler modulus.
When discussing the equation for $U_2$ to find the boundary of this inequality, we find that the solution follows the Lambert $W$ function.
From the equation of $U_2$, we can transform the expression as
\begin{align}
    - \frac{\pi}{3} U_2 e^{- \frac{\pi}{3} U_2} = - e^{- \left( A + \text{ln} \left( \frac{3}{\pi} \right) \right)}.
\end{align}
Since this relation obeys the definition of the Lambert $W$ function $W(z) e^{W(z)} = z$, the following restrictions are imposed on the imaginary part of the complex-structure modulus:
\begin{align}
    U_2 < - \frac{3}{\pi} W_{-1} \left( - \frac{\pi}{3} e^{-A} \right) = - \frac{3}{\pi} W_{-1} \left( - \frac{e^{-1 + \gamma_{\rm E}}}{4 |\eta(\omega)|^4} e^{- \frac{16 \pi^2 \text{Im} S}{b_a^{\mathcal{N} = 2}}} \right),
    \label{eq:analytical_constraint}
\end{align}
where $W_{-1}$ is the Lambert $W$ function of $-1$ branch and the modulus space is $U_2 > 1$.
Then, considering the specific value $b_a^{\mathcal{N} = 2} = 10$ and $\text{Im} S = 2$, the bound of the imaginary part of the complex-structure modulus is $U_2 \lesssim 34$.
In addition, we examine the constraints on $U_2$ with the dependence of the beta-function coefficient $b_a^{\mathcal{N} = 2}$.
The result is shown in Fig. \ref{fig:stcon}.

\begin{figure}[H]
\centering
\includegraphics[width = 0.65 \linewidth]{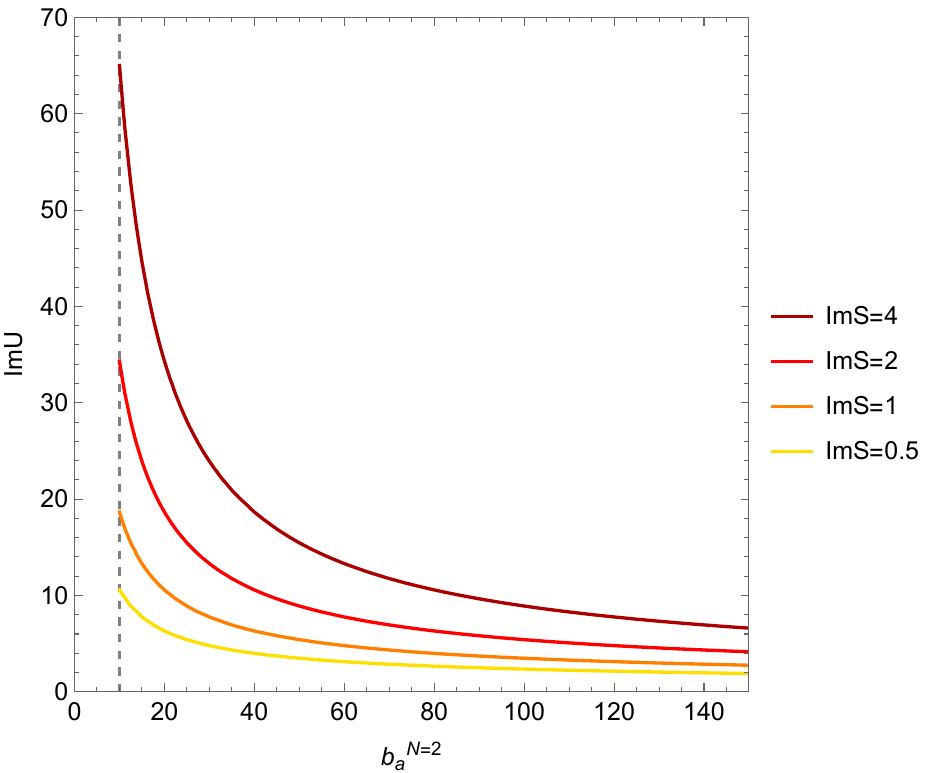}
\caption{These are the boundaries when the dilaton is varied over the range of typical values $\mathcal{O}(0.1) - \mathcal{O}(1)$ while fixing the K\"{a}hler modulus to $T = - \frac{1}{2} + i \frac{\sqrt{3}}{2}$.
Yellow, orange, red and dark red curves are respectively the boundaries with $\text{Im} S = 0.5$, $\text{Im} S = 1$, $\text{Im} S = 2$ and $\text{Im} S = 4$. The perturbative description is valid below these curves.}
\label{fig:stcon_dilaton}
\end{figure}

\noindent
Below the red region in this result, we can guarantee the perturbative theory for $\mathcal{N} = 1$ supersymmetric vacua of the heterotic string.
From this result, we can reveal that the bigger the beta-function coefficient, the stronger the bound for the imaginary part of the complex-structure modulus becomes.
Besides, we investigate the dependence of the dilaton $\text{Im} S$ in Eq. \eqref{eq:analytical_constraint}.
When the dilaton is varied over a range of typical values $\mathcal{O}(0.1) - \mathcal{O}(1)$, we analyze the change in the boundaries of the inequality Eq. \eqref{eq:analytical_constraint}.
The results are shown in Fig. \ref{fig:stcon_dilaton}.

\noindent
Above each boundary in Fig. \ref{fig:stcon_dilaton}, the perturbative description cannot be guaranteed, and it is found that the smaller the dilaton $\text{Im} S$, the stronger the constraint of the inequality.
As mentioned above, these results in Figs. \ref{fig:stcon} and \ref{fig:stcon_dilaton} are the same even when the complex-structure modulus is fixed.

\paragraph{Constraints based on $T = U$} \mbox{}

\begin{figure}[H]
\centering
\includegraphics[width = 0.65 \linewidth]{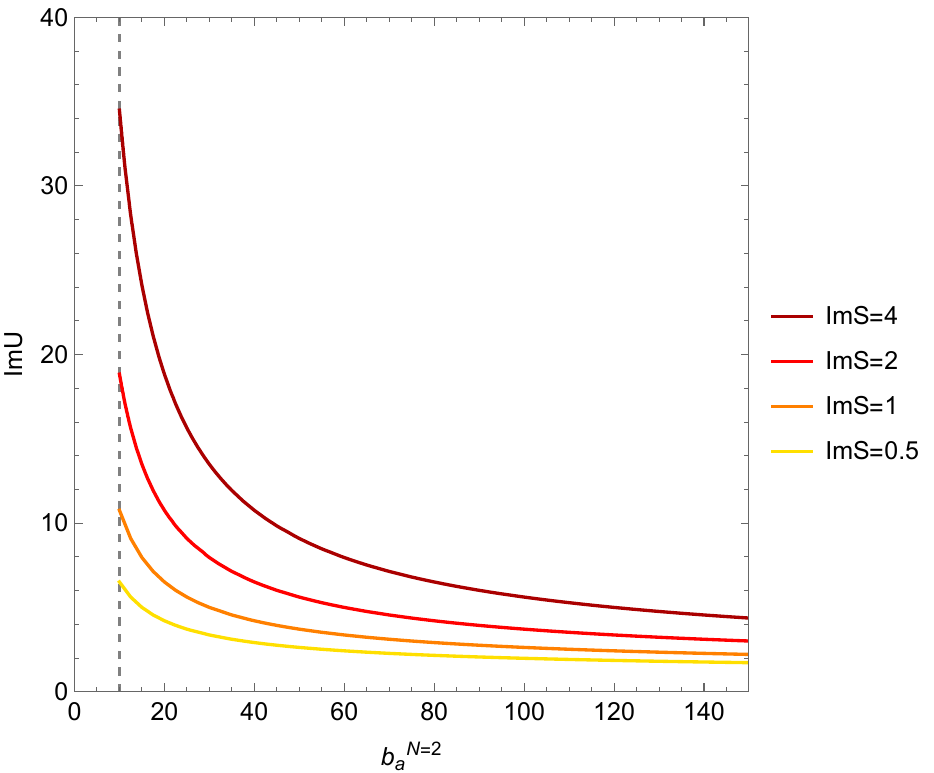}
\caption{These are the boundaries concerning $U = T$ when the dilaton is varied over the range of typical values $\mathcal{O}(0.1) - \mathcal{O}(1)$.
Yellow, orange, red and dark red curves are respectively the boundaries with $\text{Im} S = 0.5$, $\text{Im} S = 1$, $\text{Im} S = 2$ and $\text{Im} S = 4$. The perturbative description is valid below these lines.}
\label{fig:stcon_dilaton_U=T}
\end{figure}

Next, we focus on the different dependencies of the complex-structure modulus and the K\"{a}hler modulus.
Considering the case of $T = U$, Eq. \eqref{eq:analytical_constraint} can be modified as follows:
\begin{align}
    U_2 < - \frac{3}{\pi} W_{-1} \left( - \frac{\pi}{3} \left( \frac{8\pi}{3 \sqrt{3}} e^{1 - \gamma_{\rm E}} \right)^{- \frac{1}{2}} ~  ~ e^{- \frac{8 \pi^2 \text{Im} S}{b_a^{\mathcal{N} = 2}}} \right).
    \label{eq:analytical_constraintU=T}
\end{align}
We apply the same analysis to this inequality, and the results are shown in Fig.~\ref{fig:stcon_dilaton_U=T}.

\noindent
The results show that the boundaries with $U = T$ give a strong constraint on the modulus in comparison with the case where the K\"{a}hler modulus is fixed to $T = - \frac{1}{2} + i \frac{\sqrt{3}}{2}$, since the function $T_2 |\eta(T)|^4$ is maximized at $T = - \frac{1}{2} + i \frac{\sqrt{3}}{2}$.

\paragraph{Constraints based on $T = aU$} \mbox{}

\begin{figure}[H]
\centering
\includegraphics[width = 0.65 \linewidth]{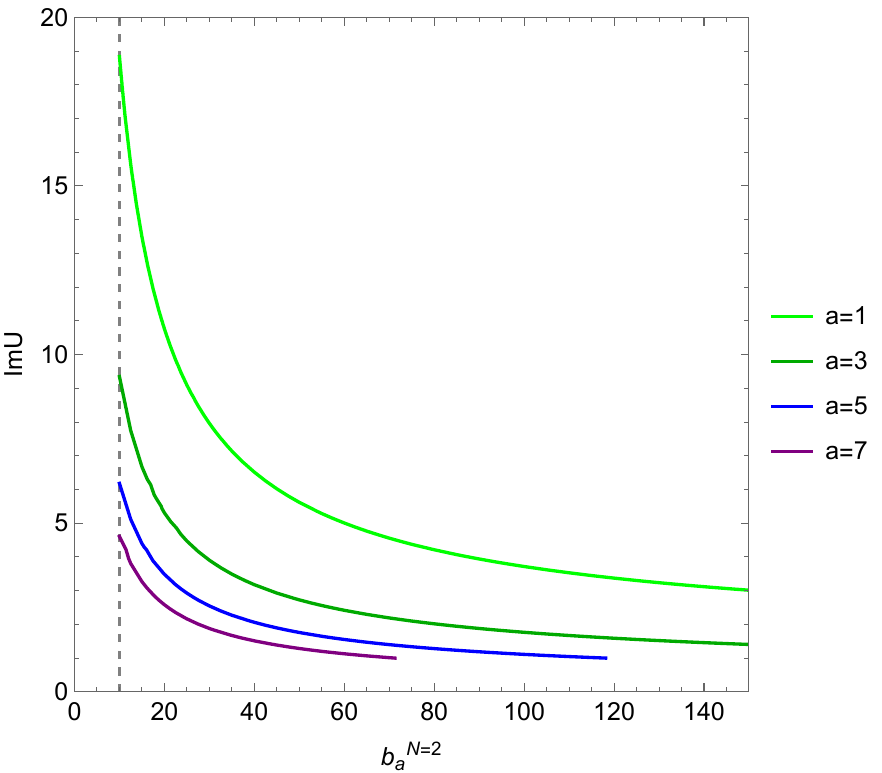}
\caption{These are the boundaries concerning $T = a U$ and $\text{Im} S = 2$ when the parameter is varied over the range from $a = 1$ to $a = 7$.
Green, dark green, blue and purple curves are respectively the boundaries with $a = 1$, $a = 3$, $a = 5$ and $a = 7$. The perturbative description is valid below these curves.}
\label{fig:stcon_T=aU}
\end{figure}

Next, we address the analysis of the stringy constraints in the case of $T = aU$.
By taking any real parameter $a$, we can discuss more general behavior of the stringy constraints.
As one can see from Eq. \eqref{eq:inequality2}, the same behavior can be obtained for $aT = U$.
Note that we have no top-down theoretical motivation to restrict $T=aU$ or $aT=U$, but that is just a working assumption.
Our moduli parameter space includes two dimensional continuous moduli.
We have discretized one parameter dimension by restricting $T=aU$ or $aT=U$ with several values of $a$.
Considering the case of $T = aU$, Eq. \eqref{eq:analytical_constraintU=T} can be modified as follows:
\begin{align}
    U_2 < - \frac{6}{(a + 1)\pi} W_{-1} \left( - \frac{(a + 1)\pi}{6 \sqrt{a}} \left( \frac{8\pi}{3 \sqrt{3}} e^{1 - \gamma_{\rm E}} \right)^{- \frac{1}{2}} ~  ~ e^{- \frac{8 \pi^2 \text{Im} S}{b_a^{\mathcal{N} = 2}}} \right).
    \label{eq:analytical_constraintT=aU}
\end{align}
Taking account of $\text{Im} S = 2$ and the range from $a = 1$ to $a = 7$, we plot the modulus as a function of the beta function coefficient in Fig.~\ref{fig:stcon_T=aU}.

\noindent
From this result, it can be seen that as the ratio of the complex-structure modulus to the K\"{a}hler modulus increases, the threshold corrections provide stronger restrictions on the modulus space than those in the case of $T = U$.
Especially, in the case of $T = 5 U$ and $T = 7 U$, $\text{Im} U$ is restricted to less than 1 within the range of large $b_a^{\mathcal{N} = 2}$.

\paragraph{Numerical constraints for moduli} \mbox{}

\begin{figure}[H]
\centering
\includegraphics[width = 0.65 \linewidth]{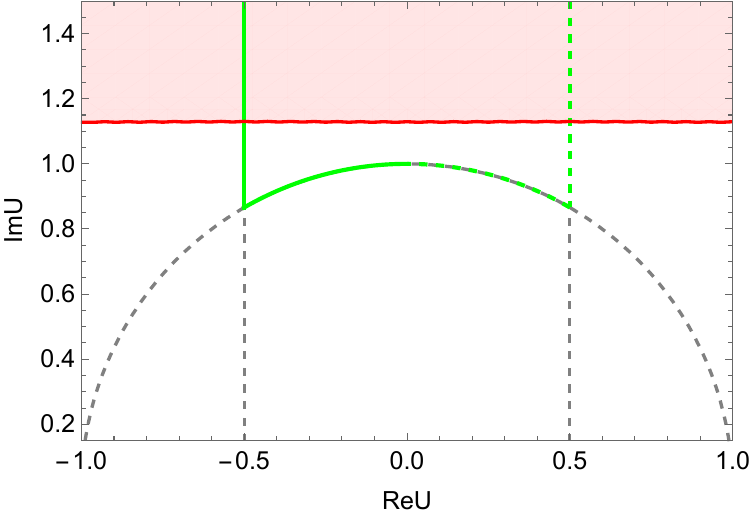}
\caption{This is the constraint concerning $T = 7 U$, $b_a^{\mathcal{N} = 2} = 60$ and $\text{Im} S = 2$.
The red region is the parameter space where the perturbative theory cannot be guaranteed.
The region framed by the green curve is the fundamental region of $PSL(2, \mathbb{Z})$.}
\label{fig:stcon_T=7U}
\end{figure}

In the previous discussion, we derive the explicit constraints on the complex-structure modulus by employing the Lambert $W$ function.
However, the function in the region $\text{Im} U \lesssim 1$ remains controversial.
To overcome this challenge, our study uses a numerical analysis based on Eq. \eqref{eq:inequality2} to explore the explicit constraints in the region.
Fig. \ref{fig:stcon_T=7U} shows the constraint with $T = 7 U$, $b_a^{\mathcal{N} = 2} = 60$ and $\text{Im} S = 2$ and the perturbative theory for $\mathcal{N} = 1$ supersymmetric vacua of the heterotic string cannot be guaranteed within the red region.

\noindent
Almost all of the fundamental region is covered with the red region, and it is found that the specific moduli spaces in this setup are incapable of discussing the perturbative theory validly.
In addition, we focus on the change of the boundaries with the different beta-function coefficients.

\begin{figure}[H]
\centering
\includegraphics[width = 0.65 \linewidth]{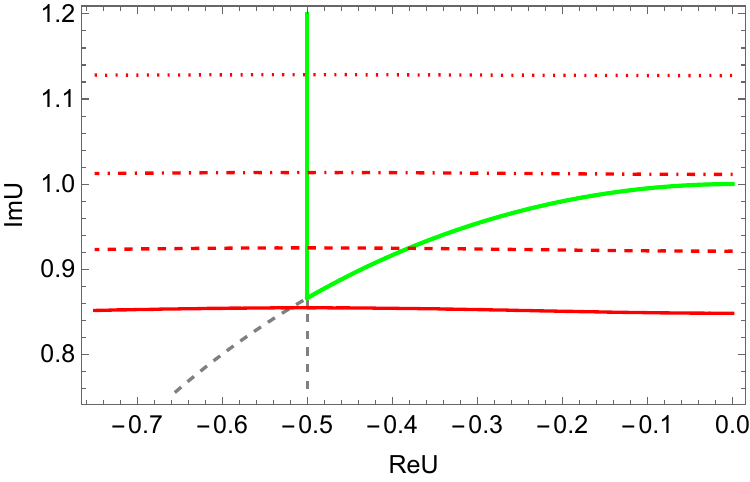}
\caption{These are the boundaries concerning $T = 7 U$ and $\text{Im} S = 2$.
Dotted red, dash-dotted red, dashed red and solid red lines are respectively the boundaries with $b_a^{\mathcal{N} = 2} = 60$, $b_a^{\mathcal{N} = 2} = 70$, $b_a^{\mathcal{N} = 2} = 80$ and $b_a^{\mathcal{N} = 2} = 90$. 
The perturbative description is valid below these lines.
The region framed by the green curve is the fundamental region of $PSL(2, \mathbb{Z})$.}
\label{fig:stcon_T=7U_anybeta}
\end{figure}

\noindent
From Fig. \ref{fig:stcon_T=7U_anybeta}, the large beta-function coefficients require strong restrictions on the moduli space, and especially the coefficient which is larger than $b_a^{\mathcal{N} = 2} = 90$ cannot guarantee the perturbative theory in the fundamental region.

%%%%%%%%%%%%%%%%%%%%%%%%%%%%%%%%%%%%%%%%%%%%%%%%%%%%%%%%%%%%
\subsection{Threshold corrections based on other modular symmetries}
\label{}
%%%%%%%%%%%%%%%%%%%%%%%%%%%%%%%%%%%%%%%%%%%%%%%%%%%%%%%%%%%%

Since the stringy constraints for the group (1) $PSL_T(2, \mathbb{Z}) \times PSL_U(2, \mathbb{Z})$ are revealed in the previous section, we investigate stringy constraints from the other target space duality symmetries (modular symmetries), presented in Sec.~\ref{sec:general}.

\paragraph{Group (2) and (3)} \mbox{}

\begin{figure}[H]
\centering
\includegraphics[width = 0.65 \linewidth]{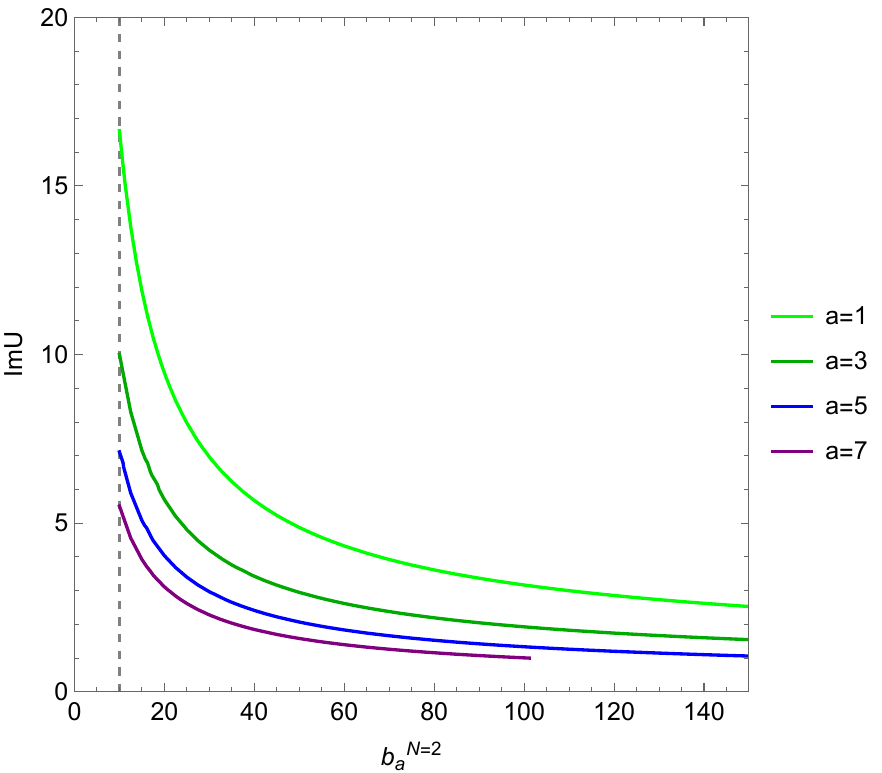}
\caption{These are the boundaries concerning $T = a U$ and $\text{Im} S = 2$ when the parameter is varied over the range from $a = 1$ to $a = 7$.
Green, dark green, blue and purple curves are respectively the boundaries with $a = 1$, $a = 3$, $a = 5$ and $a = 7$. The perturbative description is valid below these curves.}
\label{fig:stcon_group2_T=aU}
\end{figure}

\begin{figure}[H]
\centering
\includegraphics[width = 0.65 \linewidth]{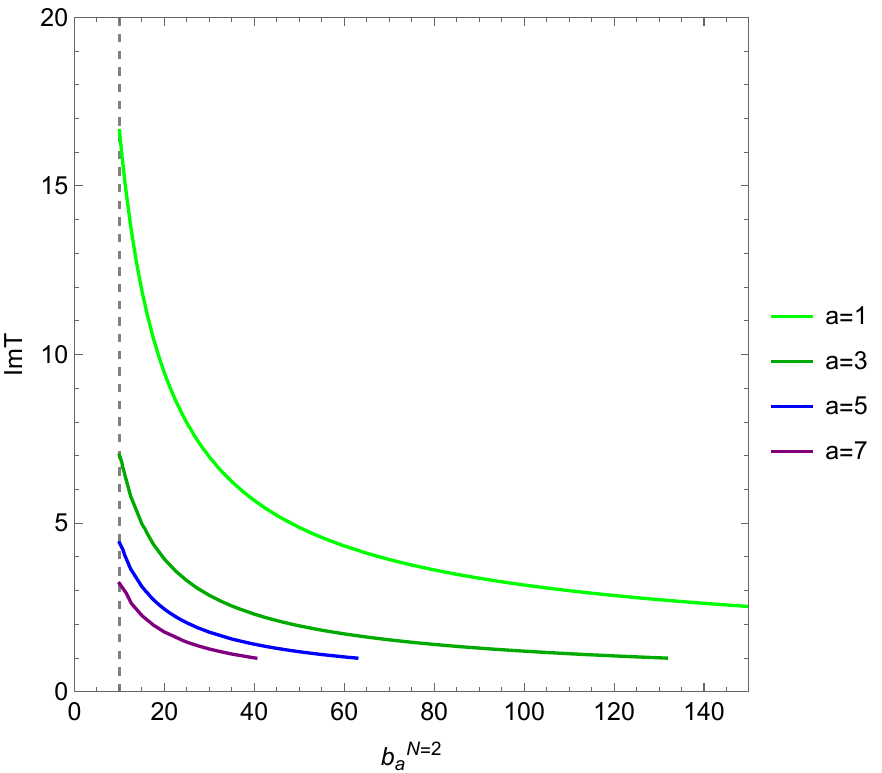}
\caption{These are the boundaries concerning $a T = U$ and $\text{Im} S = 2$ when the parameter is varied over the range from $a = 1$ to $a = 7$.
Green, dark green, blue and purple curves are respectively the boundaries with $a = 1$, $a = 3$, $a = 5$ and $a = 7$. The perturbative description is valid below these curves.}
\label{fig:stcon_group2_aT=U}
\end{figure}

Taking account of the threshold corrections of $\bar{\Gamma}_{T}^0(2) \times (\bar{\Gamma}_{U})_0(2)$ and $\bar{\Gamma}_{T}^0(2) \times PSL(2, \mathbb{Z})_U$, we construct the stringy constraints based on Eq. \eqref{eq:inequality2}.
Firstly, we focus on the case of $\bar{\Gamma}_{T}^0(2) \times (\bar{\Gamma}_{U})_0(2)$.
Since each modulus regarding $T$ and $U$ has a different dependence in the threshold corrections \eqref{eq:classifyGamma22}, it is necessary to investigate two cases of $T = aU$ and $aT = U$.
When we impose that the gauge couplings are weak enough at the string scale, the constraints for the moduli can be approximately obtained as follows:
\begin{itemize}
    \item $T = aU$
    
    \begin{align}
        U_2 < - \frac{8}{(a + 2)\pi} W_{-1} \left( - \frac{(a + 2)\pi}{8 \sqrt{a}} \left( \frac{8\pi}{3 \sqrt{3}} e^{1 - \gamma_{\rm E}} \right)^{- \frac{1}{2}} ~  ~ e^{- \frac{8 \pi^2 \text{Im} S}{b_a^{\mathcal{N} = 2}}} \right),
        \label{eq:constraint_group2_T=aU}
    \end{align}

    \item $aT = U$
    
    \begin{align}
        T_2 < - \frac{8}{(1 + 2 a)\pi} W_{-1} \left( - \frac{(1 + 2 a)\pi}{8 \sqrt{a}} \left( \frac{8\pi}{3 \sqrt{3}} e^{1 - \gamma_{\rm E}} \right)^{- \frac{1}{2}} ~  ~ e^{- \frac{8 \pi^2 \text{Im} S}{b_a^{\mathcal{N} = 2}}} \right),
        \label{eq:constraint_group2_aT=U}
    \end{align}
\end{itemize}
where we consider the region of $U_2 > 1$ and $T_2 > 1$
\footnote{
In Eq. \eqref{eq:constraint_group2_aT=U}, to examine the region where each modulus $T$ and $U$ is larger than 1 with the parameter $a$ varied over the range from $a = 1$ to $a = 7$, we describe the constraints of the K\"ahler modulus $T$.
}.
Figs.~\ref{fig:stcon_group2_T=aU} and~\ref{fig:stcon_group2_aT=U} show the constraints concerning $T = aU$ and $a T = U$ by using Eq. \eqref{eq:constraint_group2_T=aU} and Eq. \eqref{eq:constraint_group2_aT=U}.
In the case of $T = a U$ regarding $\bar{\Gamma}_{T}^0(2) \times (\bar{\Gamma}_{U})_0(2)$, the change of the constraints depending on $a$ is smaller than the case of $T = a U$ regarding $PSL_T(2, \mathbb{Z}) \times PSL_U(2, \mathbb{Z})$.
On the other hand, in the case of $aT = U$, the change of the constraints depending on $a$ is larger than the case of $PSL_T(2, \mathbb{Z}) \times PSL_U(2, \mathbb{Z})$.
It can be found that the difference between $\left| \eta \left( \frac{T}{2} \right) \right|^2$ and $|\eta(2U)|^2$ in Eq. \eqref{eq:classifyGamma22} affects the change of the constraints with the moduli dependence.

\begin{figure}[H]
 \begin{minipage}{0.49\hsize}
  \begin{center}
   \includegraphics[height=51mm]{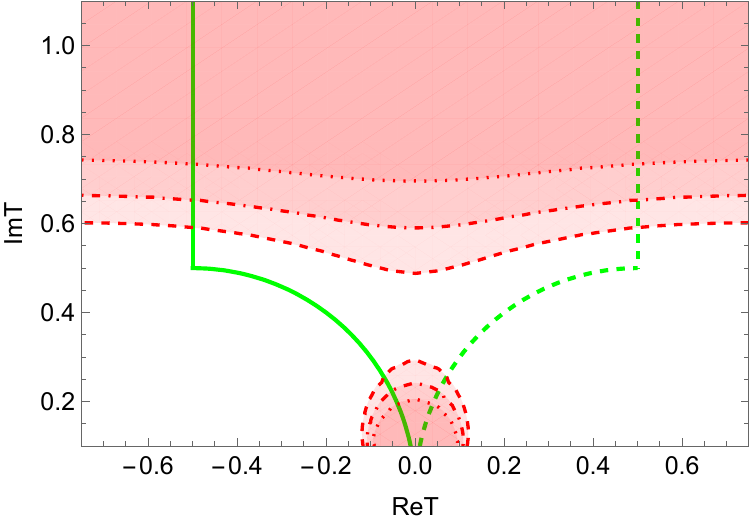}
  \end{center}
 \end{minipage}
 \begin{minipage}{0.49\hsize}
  \begin{center}
  \includegraphics[height=51mm]{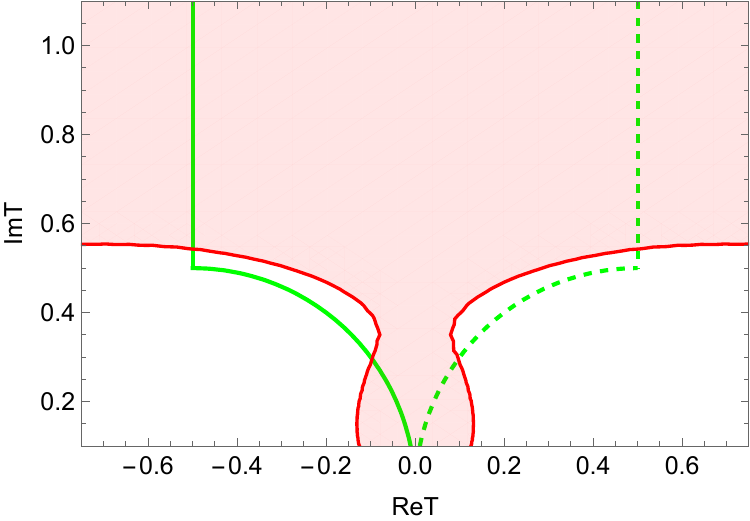}
  \end{center}
 \end{minipage}
\caption{These are the boundaries concerning $7 T = U$ and $\text{Im} S = 2$.
Dotted red, dash-dotted red, dashed red curves are respectively the boundaries with $b_a^{\mathcal{N} = 2} = 60$, $b_a^{\mathcal{N} = 2} = 70$, $b_a^{\mathcal{N} = 2} = 80$ in the left panel, and solid red curve is the boundary with $b_a^{\mathcal{N} = 2} = 90$ in the right panel. 
In both panels, the shaded regions are outside the perturbative theory, and in the left panel, when the $b_a^{\mathcal{N} = 2}$ increases, the lighter the color is. 
The region framed by the green curve is the fundamental region of $\bar{\Gamma}_0(2)$.}
\label{fig:stcon_7T=U_anybeta}
\end{figure}

As well as the analysis in the previous section, to investigate the region of $\text{Im} U \lesssim 1$ and $\text{Im} T \lesssim 1$, we perform the numerical analysis.
While the moduli space regarding $7T = U$ is restricted strongly, the fundamental region of $\bar{\Gamma}_0(2)$ is known to be enlarged from the one of $PSL(2, \mathbb{Z})$.
Therefore, it is important to explore the behavior of the numerical constraints within the region of $\text{Im} U \lesssim 1$ and $\text{Im} T \lesssim 1$ and reveal the restriction in the enlarged fundamental region.
Fig. \ref{fig:stcon_7T=U_anybeta} shows these results.

\noindent
Here, we explore the boundaries with $b_a^{\mathcal{N} = 2} = 60, 70, 80$ and $90$.
While almost all of the region concerning the upper half plane of the K\"{a}hler modulus cannot guarantee the perturbation theory in the heterotic string vacua, the enlarged fundamental region has the region regarding $\text{Im} T < \frac{\sqrt{3}}{2}$.  
Then, it is found that the region only near an elliptic point of $\bar{\Gamma}_0(2)$ ($T = - \frac{1}{2} + \frac{i}{2}$) can guarantee the perturbative theory.

For the group (3) $\bar{\Gamma}_{T}^0(2) \times PSL(2, \mathbb{Z})_U$, the constraints for the moduli can be approximately obtained by imposing that the gauge couplings are weak enough at the string scale:
\begin{itemize}
    \item $T = aU$
    
    \begin{align}
        U_2 < - \frac{12}{(a + 2)\pi} W_{-1} \left( - \frac{2^{\frac{2}{3}} (a + 2)\pi}{12 \sqrt{a}} \left( \frac{8\pi}{3 \sqrt{3}} e^{1 - \gamma_{\rm E}} \right)^{- \frac{1}{2}} ~  ~ e^{- \frac{8 \pi^2 \text{Im} S}{b_a^{\mathcal{N} = 2}}} \right),
        \label{eq:constraint_group3_T=aU}
    \end{align}

    \item $aT = U$
    
    \begin{align}
        T_2 < - \frac{12}{(1 + 2 a)\pi} W_{-1} \left( - \frac{2^{\frac{2}{3}} (1 + 2 a)\pi}{12 \sqrt{a}} \left( \frac{8\pi}{3 \sqrt{3}} e^{1 - \gamma_{\rm E}} \right)^{- \frac{1}{2}} ~  ~ e^{- \frac{8 \pi^2 \text{Im} S}{b_a^{\mathcal{N} = 2}}} \right).
        \label{eq:constraint_group3_aT=U}
    \end{align}
\end{itemize}
Since the structure of the functions is the same as the group (2) $\bar{\Gamma}_{T}^0(2) \times (\bar{\Gamma}_{U})_0(2)$ except for the coefficients, the explicit analyses are omitted in this study.

\paragraph{Group (4) and (5)} \mbox{}

\begin{figure}[H]
\centering
\includegraphics[width = 1 \linewidth]{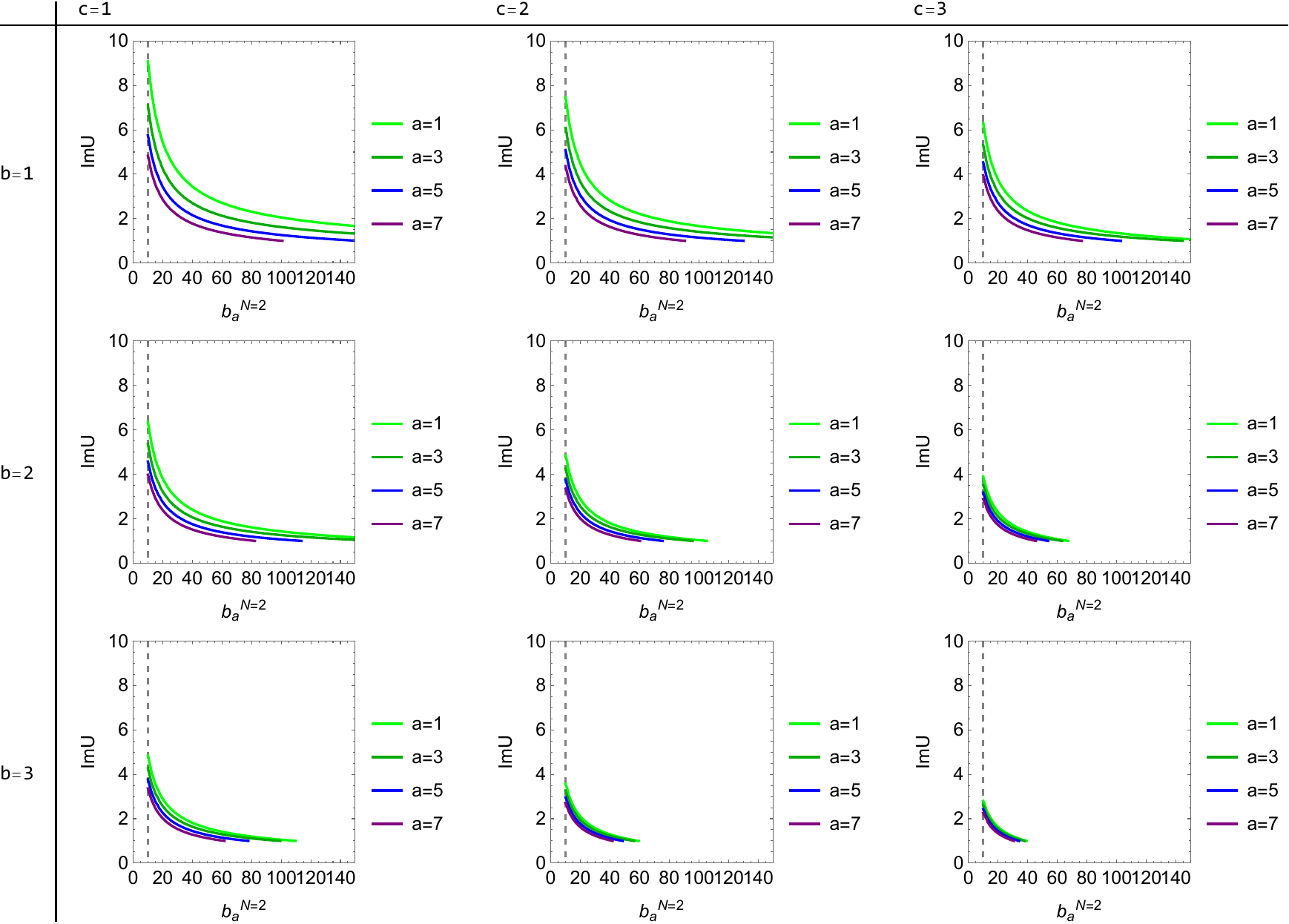}
\caption{These are the boundaries concerning $T = aU, ~ \hat{T} = bU, ~ \hat{U} = bcU$ and $\text{Im} S = 2$ when the parameter $a$ is varied over the range from $a = 1$ to $a = 7$ and the parameter $b$ and $c$ is varied over the range from $1$ to $3$.
Green, dark green, blue and purple curves are respectively the boundaries with $a = 1$, $a = 3$, $a = 5$ and $a = 7$.
In addition, we classify the cases of $b, ~ c = 1, 2, 3$. The perturbative description is valid below these curves.}
\label{fig:stcon_group4_T=aU_hatU=bU_hatT=bcU}
\end{figure}

\begin{figure}[H]
\centering
\includegraphics[width = 1 \linewidth]{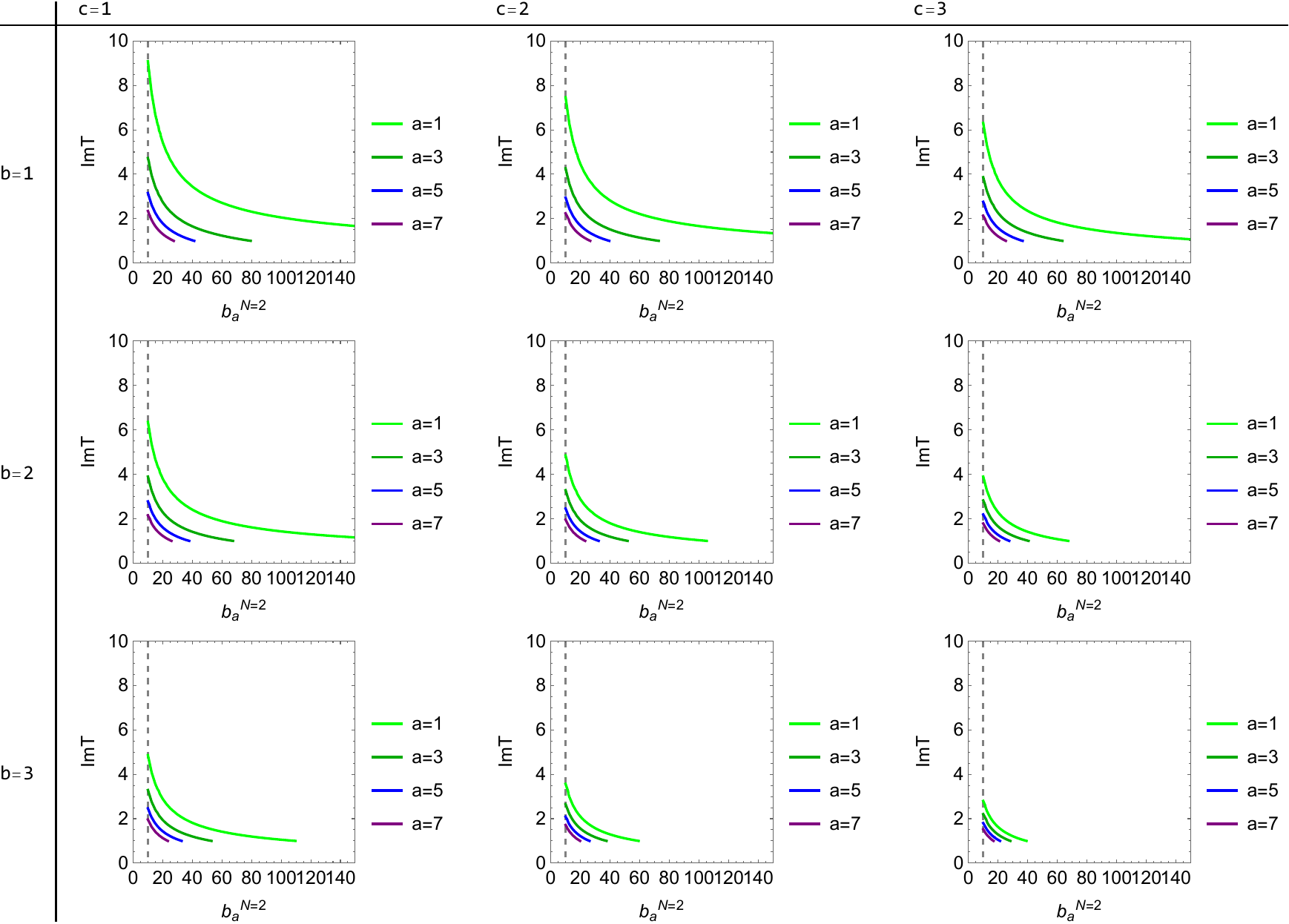}
\caption{These are the boundaries concerning $aT = U, ~ \hat{T} = bT, ~ \hat{U} = bcT$ and $\text{Im} S = 2$ when the parameter $a$ is varied over the range from $a = 1$ to $a = 7$ and the parameter $b$ and $c$ is varied over the range from $1$ to $3$.
Green, dark green, blue and purple curves are respectively the boundaries with $a = 1$, $a = 3$, $a = 5$ and $a = 7$.
In addition, we classify the cases of $b, ~ c = 1, 2, 3$. The perturbative description is valid below these curves.}
\label{fig:stcon_group4_aT=U_hatU=bU_hatT=bcU}
\end{figure}

Taking account of the threshold corrections of (4) $\bar{\Gamma}_{T}^0(3) \times (\bar{\Gamma}_{U})_0(3)$ and $PSL_{\hat{T}}(2, \mathbb{Z}) \times PSL_{\hat{U}}(2, \mathbb{Z})$, and (5) $\bar{\Gamma}_{T}^0(3) \times (\bar{\Gamma}_{U})_0(3)$ and $\bar{\Gamma}^0_{\hat{T}}(2)$, we next derive the stringy constraints based on Eq. \eqref{eq:inequality2}.
Here, we first focus on the case of the group (4).
Since each modulus regarding $T$, $U$, $\hat{T}$ and $\hat{U}$ has the different dependence in the threshold corrections \eqref{eq:classifyGamma3}, it is necessary to investigate two cases of $T = aU, ~ \hat{T} = bU, ~ \hat{U} = bcU$ and $aT = U, ~ \hat{T} = bT, ~ \hat{U} = bcT$.
The approximate constraints for the moduli can be written as follows:
\begin{itemize}
    \item $T = aU, ~ \hat{T} = bU, ~ \hat{U} = bcU$
    
    \begin{align}
        \begin{aligned}
            U_2 < - &\frac{36}{(2a + 6 + 3b (c + 1))\pi} \\
            &\times W_{-1} \left( - \frac{(2a + 6 + 3b (c + 1))\pi}{36 a^{\frac{1}{4}} b^{\frac{1}{2}} c^{\frac{1}{4}}} \left( \frac{8\pi}{3 \sqrt{3}} e^{1 - \gamma_{\rm E}} \right)^{- \frac{1}{2}} ~  ~ e^{- \frac{4 \pi^2 \text{Im} S}{b_a^{\mathcal{N} = 2}}} \right),
        \end{aligned}
        \label{eq:constraint_group4_T=aU}
    \end{align}

    \item $aT = U, ~ \hat{T} = bT, ~ \hat{U} = bcT$
    
    \begin{align}
        \begin{aligned}
            T_2 < - &\frac{36}{(2 + 6a + 3b (c + 1))\pi} \\
            &\times W_{-1} \left( - \frac{(2 + 6a + 3b (c + 1))\pi}{36 a^{\frac{1}{4}} b^{\frac{1}{2}} c^{\frac{1}{4}}} \left( \frac{8\pi}{3 \sqrt{3}} e^{1 - \gamma_{\rm E}} \right)^{- \frac{1}{2}} ~  ~ e^{- \frac{4 \pi^2 \text{Im} S}{b_a^{\mathcal{N} = 2}}} \right),
        \end{aligned}
        \label{eq:constraint_group4_aT=U}
    \end{align}
\end{itemize}
where we consider the region of $U_2 > 1$, $T_2 > 1$, $\hat{U}_2 > 1$ and $\hat{T}_2 > 1$.
The parameter $a$ plays the same role as in the previous discussion, which is associated with $T = aU$ and $aT = U$.
The parameter $b$ characterizes the difference between the moduli regarding $(I, \theta^2)$ and the moduli regarding $(I, \theta^3)$.
The parameter $c$ characterizes the difference between the K\"{a}hler modulus and the complex-structure modulus, which are based on $(I, \theta^3)$.
Similarly, in this group, $\hat{T} = bU, ~ \hat{U} = bcU$ and $\hat{T} = bcU, ~ \hat{U} = b U$ (or $\hat{T} = bT, ~ \hat{U} = bcT$ and $\hat{T} = bcT, ~ \hat{U} = bT$) provide the same results.
Figs. \ref{fig:stcon_group4_T=aU_hatU=bU_hatT=bcU} and \ref{fig:stcon_group4_aT=U_hatU=bU_hatT=bcU} show the constraints concerning $T = aU, ~ \hat{T} = bU, ~ \hat{U} = bcU$ and $aT = U, ~ \hat{T} = bT, ~ \hat{U} = bcT$ by using Eq. \eqref{eq:constraint_group4_T=aU} and Eq. \eqref{eq:constraint_group4_aT=U}.

\noindent
For this group (4) with $T = aU, ~ \hat{T} = bU, ~ \hat{U} = bcU$, it is also found that the parameter $a$ regarding the difference between $\left| \eta \left( \frac{T}{3} \right) \right|^2$ and $|\eta(3U)|^2$ in Eq. \eqref{eq:classifyGamma3} affects the change of the constraints with the moduli dependence.
When considering the parameter $b$ and $c$, it can be found that the parameter $b$ controlling the difference between the moduli of $(I, \theta^2)$ ($U, ~ T$) and the moduli of $(I, \theta^3)$ ($\hat{U}, ~ \hat{T}$) provides stronger constraints than the parameter $c$ controlling the difference between the complex-structure modulus and the K\"{a}hler modulus.

\begin{figure}[H]
\centering
\includegraphics[width = 0.65 \linewidth]{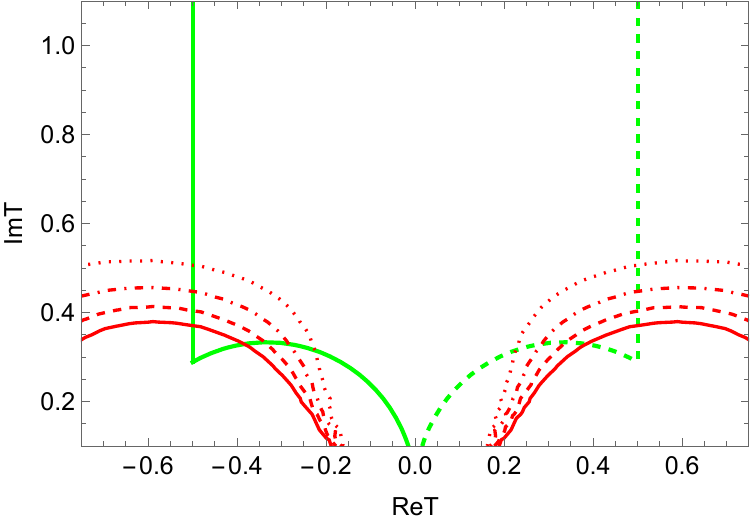}
\caption{These are the boundaries concerning $7 T = U$, $\hat{T} = T$, $\hat{U} = T$ and $\text{Im} S = 2$.
Dotted red, dash-dotted red, dashed red and solid red curves are respectively the boundaries with $b_a^{\mathcal{N} = 2} = 60$, $b_a^{\mathcal{N} = 2} = 70$, $b_a^{\mathcal{N} = 2} = 80$ and $b_a^{\mathcal{N} = 2} = 90$.
The region framed by the green curve is the fundamental region of $\bar{\Gamma}_0(3)$. The perturbative description is valid below these curves.}
\label{fig:stcon_group4_7T=U_hatU=U_hatT=U}
\end{figure}

Likewise, to investigate the region of $\text{Im} U \lesssim 1$, $\text{Im} T \lesssim 1$, $\text{Im} \hat{U} \lesssim 1$ and $\text{Im} \hat{T} \lesssim 1$, we also perform the numerical analysis in the case of the group (4).
For the congruence subgroup $\bar{\Gamma}_0(3)$, the fundamental region is known to be enlarged from the one of $\bar{\Gamma}_0(2)$.
Fig. \ref{fig:stcon_group4_7T=U_hatU=U_hatT=U} shows the restrictions in the enlarged fundamental region.

\noindent
Here, we explore the boundaries with $b_a^{\mathcal{N} = 2} = 60, 70, 80$ and $90$.
In the case of $\bar{\Gamma}_{T}^0(3) \times (\bar{\Gamma}_{U})_0(3)$ and $PSL_{\hat{T}}(2, \mathbb{Z}) \times PSL_{\hat{U}}(2, \mathbb{Z})$, it is also found that the region only near an elliptic point of $\bar{\Gamma}_0(3)$ ($T = - \frac{1}{2} + \frac{i}{2\sqrt{3}}$) can guarantee the perturbative theory.

For the group (5) $\bar{\Gamma}_{T}^0(3) \times (\bar{\Gamma}_{U})_0(3)$ and $\bar{\Gamma}^0_{\hat{T}}(2)$, the constraints for the moduli can be approximately obtained by imposing that the gauge couplings are weak enough at the string scale:
\begin{itemize}
    \item $T = aU, ~ \hat{T} = bU$
    
    \begin{align}
        \begin{aligned}
            U_2 < - &\frac{54}{(4a + 12 + 3b)}\pi \\
            &\times W_{-1} \left( - \frac{(4a + 12 + 3b)\pi}{54 a^{\frac{1}{3}} b^{\frac{1}{3}}} \left( \frac{1}{4^{\frac{2}{3}}} \frac{\sqrt{3}}{2} \left| \eta(\omega) \right|^4 \right)^{- \frac{1}{3}} \left( \frac{8\pi}{3 \sqrt{3}} e^{1 - \gamma_{\rm E}} \right)^{- \frac{2}{3}} ~  ~ e^{- \frac{16 \pi^2 \text{Im} S}{3 b_a^{\mathcal{N} = 2}}} \right),
        \end{aligned}
        \label{eq:constraint_group5_T=aU}
    \end{align}

    \item $aT = U, ~ \hat{T} = bT$
    
    \begin{align}
        \begin{aligned}
            T_2 < - &\frac{54}{(4 + 12a + 3b)}\pi \\
            &\times W_{-1} \left( - \frac{(4 + 12a + 3b)\pi}{54 a^{\frac{1}{3}} b^{\frac{1}{3}}} \left( \frac{1}{4^{\frac{2}{3}}} \frac{\sqrt{3}}{2} \left| \eta(\omega) \right|^4 \right)^{- \frac{1}{3}} \left( \frac{8\pi}{3 \sqrt{3}} e^{1 - \gamma_{\rm E}} \right)^{- \frac{2}{3}} ~  ~ e^{- \frac{16 \pi^2 \text{Im} S}{3 b_a^{\mathcal{N} = 2}}} \right).
        \end{aligned}
        \label{eq:constraint_group5_aT=U}
    \end{align}
\end{itemize}
Since the structure of the functions is the same as the group (4) $\bar{\Gamma}_{T}^0(3) \times (\bar{\Gamma}_{U})_0(3)$ and $PSL_{\hat{T}}(2, \mathbb{Z}) \times PSL_{\hat{U}}(2, \mathbb{Z})$ except for the coefficients, the explicit analyses are omitted in this study.

%%%%%%%%%%%%%%%%%%%%%%%%%%%%%%%%%%%%%%%%%%%%%%%%%%%%%%%%%%%%
\section{Phenomenological aspects}
\label{sec:phenomenological_aspects}
%%%%%%%%%%%%%%%%%%%%%%%%%%%%%%%%%%%%%%%%%%%%%%%%%%%%%%%%%%%%

In this section, we investigate the phenomenological implications of the obtained stringy constraints. 
Since the modulus value is upper bounded by a certain value, it gives a constraint on the modular form used in modular flavor models. For illustrative purposes, let us focus on the constraints for the duality group (1) $PSL_T(2, \mathbb{Z}) \times PSL_U(2, \mathbb{Z})$. 
In modular flavor models, Yukawa couplings described by modular forms have a peculiar feature around the fixed points of $PSL(2,\mathbb{Z})$ modular symmetry, i.e., $\tau=i\infty,\omega, i$. 
In the following, we analyze the phenomenological consequences of stringy constraints at three fixed points. 

\begin{figure}[H]
\centering
\includegraphics[width = 0.76 \linewidth]{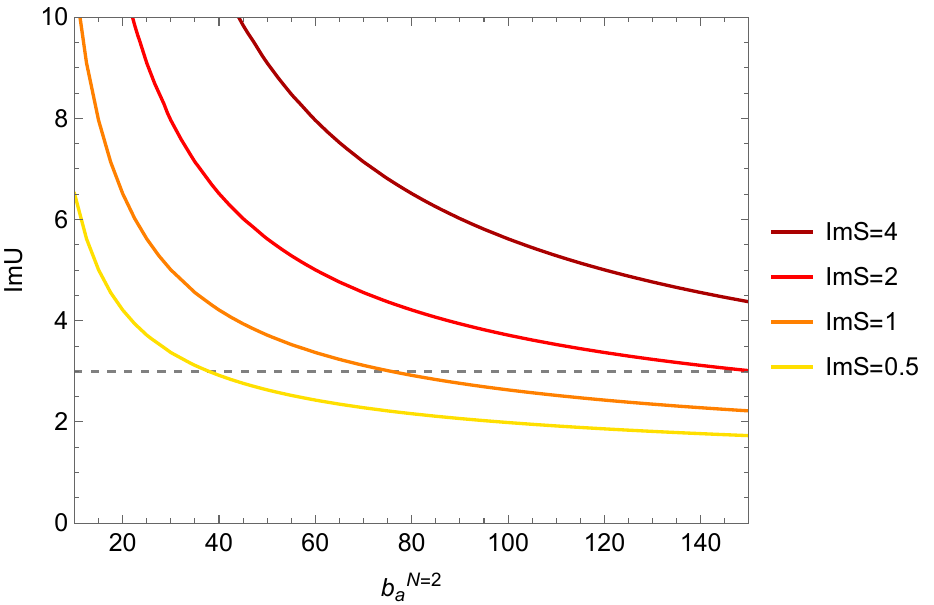}
\caption{These are the boundaries concerning $U = T$ when the dilaton is varied over the range of typical values $\mathcal{O}(0.1) - \mathcal{O}(1)$.
Yellow, orange, red and dark red curves are respectively the boundaries with $\text{Im} S = 0.5$, $\text{Im} S = 1$, $\text{Im} S = 2$ and $\text{Im} S = 4$. The dashed line describes the line of $\text{Im} U = 3$. The perturbative description is valid below these curves.}
\label{fig:stcon_dilaton_U=T_detail}
\end{figure}

Around $\tau=i\infty$, the singlet modular form of $\Gamma_N$ with weight $k$ can be expanded as
\begin{align}
\label{eq:Y_1t}
    Y_{1_t}^{(k)}(\tau)= q^{t/N}\sum_{n=0}^\infty c_n q^n\,,
\end{align}
where we denote a singlet $1_t$, corresponding to $1_0=1$, $1_1=1'$ and $1_2=1''$ for $\Gamma_3\cong A_4$. 
Since there exists the residual $\mathbb{Z}_N$ symmetry associated with $T$ transformation, one can define the charge of $1_t$ as $t$ with $0\leq t < N$ under $\mathbb{Z}_N$. 
By utilizing the structure of holomorphic modular forms \eqref{eq:Y_1t}, a hierarchical structure of 
Yukawa couplings was discussed in some modular flavor models, e.g., $\mathrm{Im}\tau\simeq 2$ for $A_4\otimes A_4\otimes A_4$ \cite{Kikuchi:2023jap},  $\mathrm{Im}\tau\simeq 2.5$ for $A_4$~\cite{Petcov:2023vws} and for $S_4'$ \cite{Abe:2023ilq,Abe:2023qmr}, $\mathrm{Im}\tau\simeq 3$ for $\Gamma_6$ \cite{Kikuchi:2023cap} and $\Gamma_6'$ \cite{Abe:2023dvr}.
When we consider the moduli space along $U=T$, a modular flavor model with large value of beta function coefficients is prohibited by stringy constraints, as shown in Fig.~\ref{fig:stcon_dilaton_U=T_detail}. 
In addition, a large value of the modulus is severely constrained by stringy constraints. 
In the radiative moduli stabilization scenario, the modulus is stabilized at the large value around ${\cal O}(10)$ in the context of modular flavor model~\cite{Kobayashi:2023spx, Higaki:2024jdk}\footnote{See, e.g., Refs.~\cite{Kobayashi:2023spx,Hai:2025wvs}, in the context of string compactifications.}, but our results indicate that the realization of such a scenario is difficult in general. 
Although a small beta function coefficient is consistent with the large value of the modulus, we often introduce extra matters to generate  Coleman-Weinberg potential to stabilize the modulus. 
Hence, it is in general too difficult to embed this scenario in stringy models. 

\begin{figure}[H]
\centering
\includegraphics[width = 0.76 \linewidth]{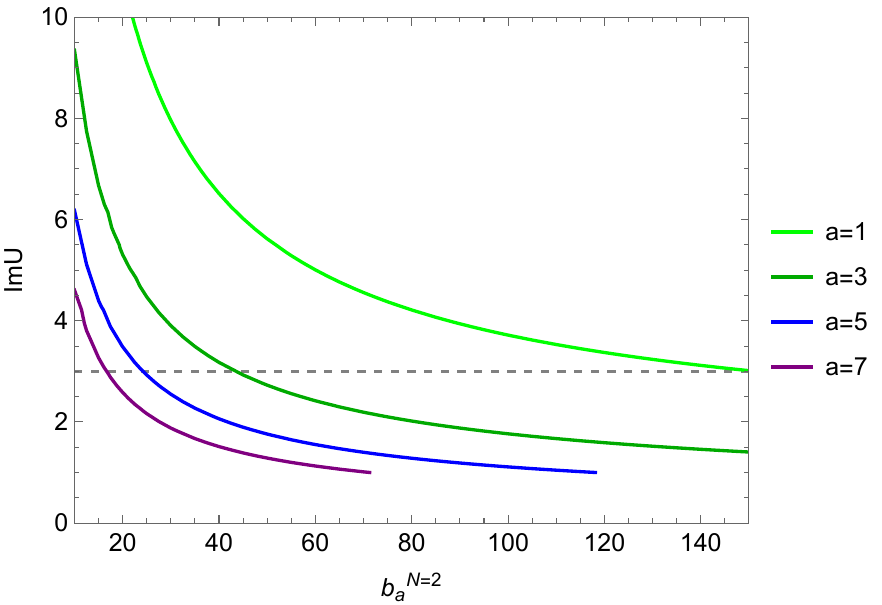}
\caption{These are the boundaries concerning $T = a U$ and $\text{Im} S = 2$ when the parameter is varied over the range from $a = 1$ to $a = 7$.
Green, dark green, blue and purple curves are respectively the boundaries with $a = 1$, $a = 3$, $a = 5$ and $a = 7$. The dashed line describes the line of $\text{Im} U = 3$. The perturbative description is valid below these curves.}
\label{fig:stcon_T=aU_detail}
\end{figure}

\begin{figure}[H]
\centering
\includegraphics[width = 0.76 \linewidth]{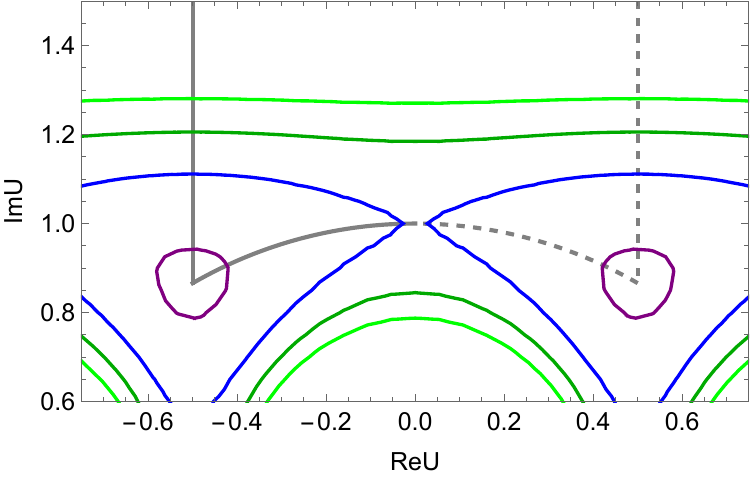}
\caption{These are the boundaries for the complex-structure modulus, which are based on $b_a^{\mathcal{N} = 2} = 60$ and $\text{Im} S = 2$. Here, the complex-structure modulus and the K\"{a}hler modulus are regarded as the independent moduli and we consider the case of the large volume $\text{Im} T > 1$.
Green, dark green, blue and purple curves are respectively the boundaries with $T = 10i$, $T = 10.02i$, $T = 10.04i$ and $T = 10.06i$. The perturbative description is valid inside of these curves.}
\label{fig:numerical_with_Kahler}
\end{figure}

Let us consider the large volume regime, i.e., $T\gg 1$ in string units, which corresponds to the parametrically controlled regime in string compactifications. 
In this regime, the modulus $U$ controlling the flavor structure of fermions is further constrained, as shown in Fig.~\ref{fig:stcon_T=aU_detail} for the case with $T=aU$ $(a=1,3,5,7)$ and Fig.~\ref{fig:numerical_with_Kahler} for $T\simeq 10i$. 
When we change the value of beta function coefficients in Fig.~\ref{fig:numerical_with_Kahler}, the constraint on $T$ is modified, e.g., $T\simeq 8i$ for $b_a^{\mathcal{N} = 2}=90$. 
Hence, the increase of $b_a^{\mathcal{N} = 2}$ gives a severe constraint on $T$.

These figures indicate that the larger the volume of extra dimensional space, the stronger the stringy constraints become. 
In particular, the $\tau=i$ region is not allowed when the volume is large enough. It has a significant impact on the modular flavor models where the flavor structure of leptons is universally predicted around $\tau=i$, as pointed out in Refs.~\cite{Feruglio:2022koo,Feruglio:2021dte}. On the other hand, the $\tau=\omega$ region still remains in the large volume limit. 
(See e.g. Refs.~\cite{Novichkov:2021evw,Petcov:2022fjf,deMedeirosVarzielas:2023crv} for phenomenological aspects around $\tau = \omega$.)
Such a fixed point is also statistically favored in a different corner of string compactifications~\cite{Ishiguro:2020tmo}, although this paper focuses on the heterotic string theory on toroidal orbifolds. 
Furthermore, our approach will be useful to phenomenological models using threshold corrections to gauge couplings. For instance, moduli-dependent threshold corrections provide a flat potential for the modulus, which will be identified as the inflaton field, as discussed in the context of string compactification~\cite{Abe:2014xja} and modular flavor models~\cite{Ding:2024neh,King_2024}.
It would be interesting to construct modular flavor models together with stringy constraints and moduli stabilization, which will be left for future work.\footnote{See, e.g., Refs.~\cite{Baur:2019kwi,Baur:2019iai,Nilles:2020nnc,Nilles:2020kgo,Nilles:2020tdp,Baur:2020jwc,Nilles:2020gvu,Ishiguro:2021drk,Ishiguro:2021ccl}, for modular flavor models in heterotic string on toroidal orbifolds.}

The above discussion is valid for the group (1), but let us comment on stringy constraints in the other groups.
The major difference between the group (1) and the other groups is the duality symmetry.
Indeed, the threshold corrections based on $\bar{\Gamma}^0(n)$ are determined by a function $\left| \eta \left( \frac{\tau}{n} \right) \right|^2$ which shifts the peak of the function in Fig. \ref{fig:shape_of_function} toward the larger imaginary part. 
As a result, the moduli space that minimizes $\Delta_a$ in Eq. \eqref{eq:inequality2} expands, and this makes it easier to guarantee perturbation theories in the broad moduli space.
On the other hand, the threshold corrections based on $\bar{\Gamma}_0(n)$ are determined by a different moduli-dependent function $\left| \eta \left( n \tau \right) \right|^2$, and the peak of the function in Fig. \ref{fig:shape_of_function} is shifted toward the smaller imaginary part.
This structure causes the moduli space, which guarantees the perturbation theories, to narrow. 
This feature can be read from Fig. \ref{fig:stcon_group2_T=aU} and Fig. \ref{fig:stcon_group2_aT=U}.
For the phenomenological aspects, unlike the group (1), it can be seen that the impact of stringy constraints on the modular flavor models depends on how the ratio of the moduli is taken, such as $T = aU$ or $aT = U$.
In particular, the ratio of $T = aU$ regarding the groups (2), (3), (4) and (5) provides the weaker stringy constraints than $T = aU$ regarding the group (1).
On the other hand, the ratio of $aT = U$ regarding the groups (2), (3), (4) and (5) provides the stronger stringy constraints than $aT = U$ regarding the group (1).

For the region of the moduli $\text{Im} U \lesssim 1$ or $\text{Im} T \lesssim 1$, we used the numerical analysis to investigate the stringy constraints due to the difficulty of the analysis in Eq.~\eqref{eq:inequality2}.
However, it would be interesting to explore the stringy constraints and the detailed structure within the regions $\text{Im} U \lesssim 1$ or $\text{Im} T \lesssim 1$ by mathematical analysis if we can generalize the Lambert $W$ function.
This will be left for future work.
Furthermore, the relations between these stringy constraints and Swampland conjectures, such as Distance conjecture and Species scale conjecture, remain an open question, and we leave a more detailed analysis for future work.

Since it is important to consider contributions related to supersymmetry breaking, we also mention these effects for the threshold corrections.
Firstly, our study has focused on supersymmetric models within the huge energy scale range from TeV to $10^{17}$ GeV, taking account of  parameter scans of the dilaton $2 \leq \text{Im} S \leq 4, ~ (25 \lesssim \alpha_a^{-1} \lesssim 50)$.
Since the supersymmetry breaking in the MSSM at the TeV scale can be evaluated as little effect for the threshold corrections, we focus on the high-scale MSSM ($\text{Im}S = 4, (\alpha_a^{-1} \sim 50)$).
Here, considering that the unification of couplings occurs at a Kaluza-Klein scale $M_{\rm KK}$, we evaluate a simple example where the threshold corrections arise mainly from massive KK modes.
In addition, we focus on the threshold corrections in ordinary gauge theories \cite{kaplunovsky:1992}.
Then, the leading term of the difference $\delta \Delta_a$, which originates from supersymmetric breaking $M_n^2 \to M_n^2 + m_{\rm soft}^2$, can be expressed by
\begin{align}
    \begin{aligned}
        \delta \Delta_a &\sim \sum
        \ln \left(\frac{M_n^2+m_{\rm soft}^2}{M_n^2}  \right) \\
        & \sim 
        \left( \frac{m_{\rm soft}}{M_{\rm KK}} \right)^2 \sum_{n=1}^\infty \frac{1}{n^2} \\
        & \sim  \left( \frac{m_{\rm soft}}{M_{\rm KK}} \right)^2,
    \end{aligned}
\end{align}
%\begin{align}
%    \delta \Delta_a \sim \frac{1}{3} T_a (R) \left( \frac{m_{\rm soft}}{M_{\rm KK}} \right)^2 \sum_{n=1}^\infty \frac{1}{n^2} = \frac{\pi^2}{18} T_a (R) \left( \frac{m_{\rm soft}}{M_{\rm KK}} \right)^2,
%\end{align}
where we have assumed the mass of the $n$-th KK modes $M_n = n M_{\rm KK}$.
%Here, 
%$T_a (R)$ is a Dynkin index of representation $R$.
%Note that we use the relation of $\sum_n 1/n^2 = \pi^2/6$.
%Even by a conservative estimate, $\frac{\pi^2}{18} T_a (R)$ is considered to be roughly $\mathcal{O} (1)$ or less.
When considering the mass scale characterizing the supersymmetry breaking, which is $m_{\rm soft} \sim M_{\rm KK}/10$, the difference of the threshold corrections can be evaluated roughly as $\delta \Delta_a \lesssim \mathcal{O} (10^{-2})$.
Based on the most conservative estimates, the smallest value of the full threshold correction originating from string theory in Eq. \eqref{eq:RHS} is approximately 0.680.
Thus, the ratio between the threshold correction and the difference is found to be $\delta \Delta_a/\Delta_a \lesssim 1.5\%$.
Therefore, as long as we consider $m_{\rm soft} \lesssim M_{\rm KK}/10$ as the upper limit of the scale regarding supersymmetry breaking, we can conclude that the validity is guaranteed.
Moreover, if we consider models whose scale regarding supersymmetry breaking is higher than the KK scale or models with respect to non-SUSY string theory, we must address more robust evaluations of the threshold corrections and the differences concerning the supersymmetry breaking.
We would like to explore this in our future research.

%%%%%%%%%%%%%%%%%%%%%%%%%%%%%%%%%%%%%%%%%%%%%%%%%%%%%%%%%%%%
\section{Conclusions}
\label{sec:conclusion}
%%%%%%%%%%%%%%%%%%%%%%%%%%%%%%%%%%%%%%%%%%%%%%%%%%%%%%%%%%%%

In this work, we used the moduli-dependent threshold corrections in heterotic string vacua to investigate the stringy constraints on the moduli spaces in the modular flavor models. 
As discussed in Ref. \cite{BAILIN:1994}, the combinations of the duality symmetries $PSL(2, \mathbb{Z})$, $\bar{\Gamma}_0(n)$ and $\bar{\Gamma}^0(n)$ can classify the characteristics of the threshold corrections on each toroidal orbifold.

To validate the perturbative description of $\mathcal{N} = 1$ supersymmetric heterotic string vacua, the tree-level gauge coupling determined by the dilaton must be larger than the moduli-dependent threshold corrections, and the inequality with respect to the complex-structure modulus $U$ and the K\"{a}hler modulus $T$ can be obtained.
When exploring the inequality in the region where $\text{Im} U$ and $\text{Im} T$ are large enough, it was approximately found that the boundary as a function of $U$ and $T$ follows the Lambert $W$ function.
Our study used this function to reveal the behavior of the stringy constraints that have the dependence of the 4D dilaton $\text{Im} S \sim \mathcal{O}(1)$, the beta-function coefficient $|b_a^{\mathcal{N} = 2}| \sim \mathcal{O}(10)$ and the ratio between $T$ and $U$, where our study focused on only $k_a  = 1$ cases. 
Then, our finding results systematically showed that (1) the smaller $\text{Im} S$, the stronger the stringy constraints, (2) the larger $b_a^{\mathcal{N} = 2}$, the stronger the stringy constraints, and (3) the larger the ratio between $T$ and $U$, the stronger the stringy constraints.
In particular, Figs.~\ref{fig:stcon_7T=U_anybeta} and \ref{fig:stcon_group4_7T=U_hatU=U_hatT=U} indicate that the moduli space around the fixed point of the duality symmetries seems to be good candidates for the region which are allowed in the context of the perturbative theories. In this paper, we have focused on toroidal orbifolds, but it is interesting to explore other background geometries such as Calabi-Yau threefolds, which will be left for future work.

For the phenomenological aspects, the obtained stringy constraints put a severe constraint on the moduli values in the modular flavor model and the radiative moduli stabilization. 
Basically, the constraints cannot allow the large imaginary parts of the moduli $\tau = i \infty$, as shown in Refs.~\cite{ANTONIADIS1990377, IBANEZ1991251, BARBIERI199479, FLORAKIS20171, ANGELANTONJ2019496}, and the modular flavor models which are necessary to choose the large $\text{Im} \tau$ and the large beta-function coefficients remain controversial.
In the large volume regime, i.e., $T\gg 1$, the allowed region of the complex-structure modulus is severely restricted.
When considering the large volume limit, it was found that the fixed point $\tau = i$ is more restricted than the fixed point $\tau = \omega$.
This difference of the fixed points offers interesting implications for the phenomenologically viable modular flavor models.
Moreover, the modular flavor models and the modular inflation models, which are based on the stringy constraints of the threshold corrections, need further investigation.
This work is restricted to a perturbative analysis and non-perturbative effects are not included.
Such effects might modify our conclusions. 
A systematic treatment incorporating non-perturbative contributions is left for future work.

\acknowledgments

This work was supported in part by K2-SPRING program Grant number JPMJSP2136 (T. Kai) and JSPS KAKENHI Grant Numbers JP23K03375 (T. Kobayashi) and JP25H01539 (H.O.).

All views and conclusions expressed in this paper are solely those of the authors and should not be attributed to SMBC Nikko Securities Inc.

\bibliography{references}{}
\bibliographystyle{JHEP}

\end{document}